\documentclass[conference,a4paper]{IEEEtran}

\pdfpageattr{/Group << /S /Transparency /I true /CS /DeviceRGB >>}
\usepackage[utf8]{inputenc}
\usepackage[T1]{fontenc}
\usepackage{textcomp}         % load ONCE
\usepackage{microtype}
\usepackage{calc}
\usepackage{xspace}

\usepackage{silence}
\usepackage{amsmath}
\usepackage{amssymb}
\usepackage{amsfonts}
\usepackage{amsthm}

\usepackage{booktabs}
\usepackage{array}
\usepackage{multirow}

\usepackage{algorithm}
\usepackage{algpseudocode}

\usepackage[
  detect-all,
  forbid-literal-units,
  range-phrase=--,
  per-mode=symbol-or-fraction,
  range-units=single,
  list-units=single
]{siunitx}

\AtBeginDocument{
  \DeclareSIUnit\byte{Byte}
  \DeclareSIUnit\decibeli{dBi}
  \DeclareSIUnit\decibelm{dBm}
  \DeclareSIUnit\megasamplespersecond{Msps}
  \DeclareSIUnit\dbm{dBm}
  \DeclareSIUnit\ppm{ppm}
  \DeclareSIUnit\watthour{Wh}
  \DeclareSIUnit\herz{Hz}
  \sisetup{output-exponent-marker=\textsc{e}}
}

\usepackage{xcolor}
\usepackage{listings}
\usepackage{graphicx}
\DeclareGraphicsExtensions{.pdf,.png,.jpg}

\usepackage[caption=false,font=footnotesize]{subfig}
\usepackage{tikz}
\usetikzlibrary{calc}
\usepackage{pgfplots}
\pgfplotsset{compat=newest}
\pgfplotsset{plot coordinates/math parser=false}

\usepackage{pifont}
\newcommand{\cmark}{\ding{51}} % ✓
\newcommand{\xmark}{\ding{55}} % ✗

\usepackage{verbatim}

\usepackage[american]{babel}
\usepackage{hyphenat}
\RequirePackage{xstring}
\RequirePackage{xparse}
\RequirePackage{acro}
\makeatletter
\@ifpackagelater{acro}{2019/02/17}{
  \NewDocumentCommand\acrodef{mO{#1}mG{}}{\DeclareAcronym{#1}{short={#2}, long={#3}, foreign-plural={}, #4}}
}{
  \NewDocumentCommand\acrodef{mO{#1}mG{}}{\DeclareAcronym{#1}{short={#2}, long={#3}, #4}}
}
\makeatother

\usepackage{csquotes}
\usepackage[
  backend=biber,
  style=ieee,
  doi=false,
  isbn=false,
  maxnames=99,
  mincitenames=1,
  maxcitenames=2
]{biblatex}
\DeclareFieldFormat{sentencecase}{#1}
\DeclareFieldFormat{titlecase}{#1}
\usepackage{xpatch}
\xpatchbibmacro{textcite}{\addspace}{\addnbspace}{}{}
\xpatchbibmacro{Textcite}{\addspace}{\addnbspace}{}{}
\DefineBibliographyStrings{english}{andothers = et~al\adddot\addspace}

\DeclareSourcemap{
  \maps[datatype=bibtex]{
    \map[overwrite=true]{
      \step[fieldsource=school,fieldtarget=institution]
    }
  }
}

\usepackage{url}
\usepackage{hyperref}
\usepackage[capitalize,noabbrev,nameinlink]{cleveref}

\crefname{figure}{Fig.}{Figs.}
\crefname{table}{Table}{Tables}
\crefname{section}{Section}{Sections}

\usepackage{todonotes}
\usepackage[inline]{enumitem}

\acrodef{Adam}{adaptive moment estimation}
\acrodef{AoI}{age of information}
\acrodef{AKF}{adaptive Kalman filter}
\acrodef{AltDS-TWR}{alternative double-sided two-way ranging}
\acrodef{AoA}{angle of arrival}
\acrodef{AT-LSTM}{attention-based LSTM}
\acrodef{Bi-LSTM}{bidirectional LSTM}
\acrodef{BERT}{bidirectional encoder representations from transformers}
\acrodef{CDF}{cumulative distribution function}
\acrodef{CIR}{channel impulse response}
\acrodef{CNN}{convolutional neural network}
\acrodef{DEKF}{double extended Kalman filter}
\acrodef{DGAN}{diffusion-based generative adversarial network}
\acrodef{DL}{deep learning}
\acrodef{DR}{dead reckoning}
\acrodef{DWIFCM}{dynamic weight integrated fuzzy C-means}
\acrodef{6-DOF}{six-degree-of-freedom}
\acrodef{EKF}{extended Kalman filter}
\acrodef{ENU}{east-north-up}
\acrodef{GAN}{generative adversarial network}
\acrodef{GDOP}{geometric dilution of precision}
\acrodef{GNSS}{global navigation satellite system}
\acrodef{GRU}{gated recurrent unit}
\acrodef{IMU}{inertial measurement unit}
\acrodef{KF}{Kalman filter}
\acrodef{LOS}{line-of-sight}
\acrodef{Bi}{Bidirectional}
\acrodef{LSTM}{long short-term memory}
\acrodef{MAE}{mean absolute error}
\acrodef{MEMS}{microelectromechanical system}
\acrodef{MLE}{maximum likelihood estimation}
\acrodef{MLP}{multi-layer perceptron}
\acrodef{MSE}{mean squared error}
\acrodef{NLOS}{non-line-of-sight}
\acrodef{PDR}{pedestrian dead-reckoning }
\acrodef{PF}{particle filter}
\acrodef{PSO}{particle swarm optimization}
\acrodef{RF}{radio frequency}
\acrodef{RMSE}{root mean square error}
\acrodef{RSSI}{received signal strength indicator}
\acrodef{RTK}{real-time kinematic}
\acrodef{SO}{special orthogonal group}
\acrodef{SVM}{support vector machine}
\acrodef{TDoA}{time difference of arrival}
\acrodef{TL}{transfer learning}
\acrodef{ToA}{time of arrival}
\acrodef{ToF}{time of flight}
\acrodef{TWR}{two-way ranging}
\acrodef{UAV}{unmanned aerial vehicle}
\acrodef{UFIR}{unbiased finite impulse response}
\acrodef{UKF}{unscented Kalman filter}
\acrodef{UWB}{ultra-wideband}
\acrodef{WBO}{weight-based optimization}
\acrodef{WCE}{weighted centroid estimation}
\acrodef{WMSE}{weighted mean squared-error}
\acrodef{WSN}{wireless sensor networks}
\acrodef{XGBoost}{extreme gradient boosting}

\usepackage[detect-all,forbid-literal-units,per-mode=symbol]{siunitx} %binary-units
\DeclareSIUnit\byte{Byte}
\DeclareSIUnit\decibeli{dBi}
\DeclareSIUnit\decibelm{dBm}
\DeclareSIUnit\megasamplespersecond{Msps}
\DeclareSIUnit\dbm{dBm}
\DeclareSIUnit\ppm{ppm}
\DeclareSIUnit\watthour{Wh}
\DeclareSIUnit\herz{Hz}
\DeclareSIUnit\byte{Bytes}
\newcommand{\proposalName}{\texttt{AoI-FusionNet}}
\newcommand{\DGAN}{\texttt{AoI-FusionNet-DGAN}}
\IEEEoverridecommandlockouts

\begin{document}

\title{Age-Aware Tightly Coupled Fusion of UWB-IMU under Sparse Ranging Conditions}
\title{AoI-FusionNet: Age-Aware Tightly Coupled Fusion of UWB-IMU under Sparse Ranging Conditions}

% \author{%
%     Tehmina Bibi, \IEEEmembership{Student Member, IEEE},
%     Anselm Köhler, %\IEEEmembership{Senior Member, IEEE},
%     Jan-Thomas Fischer, %\IEEEmembership{Student Member, IEEE},
%     Falko Dressler,  \IEEEmembership{Fellow, IEEE}
%     \thanks{T. Bibi and F. Dressler are with School of Electrical Engineering and Computer Science, TU Berlin, Germany (e-mail: \{bibi,dressler\}@ccs-labs.org).}
%     \thanks{A. Köhler and J.-T. Fischer are with the Austrian Research Centre for Forests, Austria (e-mail: \{anselm.koehler,jt.fischer\}@bfw.gv.at).}
%     \thanks{This work was supported by the project AvaRange funded by German Research Foundation (DFG), grant DR 639/22-1 and Austrian Science Fund (FWF), grant I 4274-N29.}
% }

\author{
    \IEEEauthorblockN{
    Tehmina Bibi\IEEEauthorrefmark{1},
    Anselm Köhler\IEEEauthorrefmark{2},
    Jan-Thomas Fischer\IEEEauthorrefmark{2},
    and Falko Dressler\IEEEauthorrefmark{1}
    }
    \IEEEauthorblockA{\IEEEauthorrefmark{1}School of Electrical Engineering and Computer Science, TU Berlin, Germany}
    \IEEEauthorblockA{\IEEEauthorrefmark{2}Austrian Research Centre for Forests, Austria}
    \texttt{\{bibi,dressler\}@tkn.tu-berlin.de}, \texttt{\{anselm.koehler,jt.fischer\}@bfw.gv.at}
    \thanks{This work was supported by the project AvaRange funded by German Research Foundation (DFG), grant DR 639/22-1 and Austrian Science Fund (FWF), grant I 4274-N29.}
}

\maketitle 
\begin{abstract}
Accurate motion tracking of snow particles in avalanche events requires robust localization in \ac{GNSS}-denied outdoor environments.
%, which remains challenging due to sparse anchor visibility, intermittent ranging, and \ac{NLOS} effects.
%
This paper introduces \proposalName{}, a tightly coupled deep learning based fusion framework that directly combines raw \ac{UWB} \ac{ToF} measurements with \ac{IMU} data for 3D trajectory estimation.
Unlike loose-coupled pipelines based on intermediate trilateration, the proposed approach operates directly on heterogeneous sensor inputs, thereby, enabling localization even under insufficient ranging availability.
This framework integrates an \ac{AoI}-aware decay module to reduce the influence of stale \ac{UWB} rangings and a learned attention gate mechanism that adaptively balances trust between \ac{UWB} and \ac{IMU}
modalities as a function of measurement availability and temporal freshness.
To evaluate robustness under limited data and measurement variability, we apply a diffusion-based residual augmentation strategy %(\ac{DGAN})
during training, producing an augmented variant termed \DGAN{}.
We assess the performance of the proposed model using offline post-processing of real-world measurement data from an alpine environment and benchmark against \ac{UWB} multilateration and loose-coupled fusion baselines. % such as \ac{AKF} and \ac{Bi-LSTM}.
%
% AoI-FusionNet %
% % and AoI-FusionNet-DGAN 
% consistently achieves lower \ac{RMSE} and tail errors than all baselines. % especially AoI-FusionNet-DGAN reaches an \ac{RMSE} of \SI{0.129}{\meter} with 95th and 99th percentile errors dropped to \SI{0.259}{\meter} and \SI{0.355}{\meter} respectively.
Our results demonstrate that \proposalName{} substantially reduces mean and tail localization errors under intermittent and degraded sensing conditions.
\end{abstract}

\begin{IEEEkeywords}
 Ultra-wideband (UWB), inertial measurement unit (IMU), tightly coupled fusion, deep learning
\end{IEEEkeywords}

\maketitle

\acresetall

\section{Introduction}\label{sec:intro}

Motion tracking on a particle level inside extreme natural flow phenomena such as snow avalanches~\cite{neuhauser2023particle} is particularly challenging, as dynamic conditions, limited visibility of satellites, and poor geometric configurations can lead to reduced localization accuracy. 
Harsh terrain, signal attenuation, chaotic motion, and significant snow depth severely limit the applicability of \ac{GNSS} and make real-time localization using single-sensor unreliable~\cite{wang2024dynamic}. 
Also, size and energy constraints of high-precision \ac{GNSS} systems are often prohibitive for use in mobile devices.

\Ac{UWB} has emerged as a very promising alternative in \ac{GNSS}-denied environments since it uses short pulses, wider bandwidth, and achieves centimeter-level ranging accuracy~\cite{arslan2006ultra}. However, \ac{NLOS} conditions and multipath effects affect \ac{UWB}, ultimately leading to limited connectivity and intermittent rangings.
Consumer-grade \acp{IMU}, in contrast, deliver high frequency motion data enabling accurate short-term inertial dead-reckoning but suffer from accumulation of drift over time~\cite{yang2021novel} and therefore are unable to maintain long-term precision.
To achieve accurate positioning under these conditions, a fusion of complementary sensing modalities is required.
% Consequently, single-sensor-based localization techniques exhibit inherent drawbacks, motivating sensor fusion where complementary sensing modalities offset each other's weaknesses to achieve reliable and accurate tracking.
%
Existing fusion techniques typically adopt a loosely coupled architecture that first processes \ac{UWB}-based \ac{ToF} measurements using trilateration to generate position estimates to later fuse with inertial data~\cite{cheng2024uwbins} via Bayesian filtering or learning-based fusion models~\cite{zhang2019outages, chen2024error}. 
%
% Such methods are effective under favorable conditions but discard important information during preprocessing when anchor availability is insufficient for trilateration.
% In practice, it is common in avalanche scenarios, with sparse rangings being classified as outliers when insufficient \ac{ToF} data are available for trilateration.
%
% Data-driven fusion approaches based on deep learning have recently been extensively explored by researchers due to their capability to learn complex nonlinear relationships directly from sensor data, especially when measurements deviate from Gaussian noise assumption.
% %
% For \ac{UWB} \ac{LOS} and \ac{NLOS} conditions, the learning methods range from traditional machine-learning methods such as decision tree\cite{zhu2023robust} to \ac{DL}, including \ac{CNN}~\cite{jiang2020UWB}.
% %
% Earlier methods using \ac{MLP}\cite{si2023lightweight} and \ac{SVM} \cite{cong2020vector} provided limited capability for temporal modeling.
% %
% These shortcomings have been addressed by recurrent architectures such as \ac{LSTM} networks\cite{kim2023positioning}, \ac{GRU}\cite{zhang2019outages,li2020GRU}, and attention-based models \cite{chen2024AT}, which are well suited for capturing long-term temporal dependencies.

In the scope of the AvaRange project, we developed an in-flow sensor system in which wireless sensor nodes are embedded in snow for data acquisition during avalanches~\cite{kuss2023measurement}.
To track individual particles within the flow, a sensor network is created by placing static anchors (called AvaAnchors) across a large alpine area.
Snow particle-shaped, 3D printed mobile sensor nodes (called AvaNode) are equipped with a multi-sensor setup where \ac{ToF} ranging is performed by \ac{UWB} sensor, and an \ac{IMU} sensor produces motion data.
The position estimation of the mobile node is performed by measuring distances to the static anchors.
The tracking results from this experiment confirmed that \ac{UWB}-based ranging using \ac{ToF} measurements has better accuracy than conventional localization methods such as \ac{GNSS} \cite{kuss2024distributed}. 
Unfortunately, \ac{LOS} conditions were not always achievable, and localization accuracy is significantly affected by multipath effects, \ac{NLOS} conditions, or \ac{LOS} signal blockage.
This leads to intermittent \ac{LOS}/\ac{NLOS} ranging conditions.

Existing mitigation and correction strategies either depend on dense anchor deployments, computationally expensive post-processing algorithms~\cite{xiao2014mitigation,zhang2019direction,zhang2013kurtosis}, or identification of \ac{NLOS} measurements and their subsequent classification as outliers by excluding them using localization filters~\cite{dwek2020improving}. 
Classical tightly coupled \ac{KF}-based fusion methods~\cite{zhu2019bias,feng2020kalman,ali2021tightly} rely on recursive state estimates, making them sensitive to model inaccuracies and difficult to apply in nonlinear or non-Gaussian scenarios \cite{hashim2018stochastic}.
More recent learning-based ranging compensation methods mitigate \ac{NLOS} bias without discarding measurements; however, they still assume a sufficient number of simultaneous ranges for trilateration and therefore remain vulnerable under sparse anchor visibility or intermittent ranging conditions~\cite{shalihan2025localization}.
Together, this highlights the need for a fusion framework that directly processes raw measurements and models their temporal reliability.

Motivated by these challenges, we developed a data-driven, tightly coupled \ac{UWB}-\ac{IMU} fusion framework that performs joint temporal reasoning over heterogeneous sensor measurements and remains operational even when an insufficient number of \ac{UWB} measurements are available to produce a standalone positioning.
In particular, we propose {\proposalName}, a deep-learning-based fusion architecture that directly integrates raw \ac{UWB} range measurements with \ac{IMU} acceleration measurements without relying on intermediate trilateration.
Under sparse and intermittent ranging conditions, the temporal freshness of \ac{UWB} measurements plays a role comparable to its availability.
To incorporate this, {\proposalName} uses an \ac{AoI}-aware decay mechanism that down-weights stale \ac{UWB} rangings, combined with a learned attention gate that adaptively balances the relative trust between \ac{UWB} and \ac{IMU} measurements. 
The proposed model learns to predict incremental 3D motion displacements over a sliding temporal window and reconstructs the full trajectory through accumulation, thus improving stability and long-term horizon drift.
To mitigate overfitting due to data scarcity, we introduced a data augmentation strategy that injects physically plausible \ac{UWB} perturbations during training to improve stress-testing of robustness without changing the underlying motion trajectory. 

Our main contributions can be summarized as follows:
\begin{itemize}

\item We propose {\proposalName}, a tightly coupled deep learning-based sensor fusion framework that fuses raw \ac{UWB} range measurements with time-synchronized \ac{IMU} acceleration data without relying on trilateration.

\item We introduce an \ac{AoI}-aware decay mechanism and a learned attention gate that dynamically modulate the influence of \ac{UWB} and \ac{IMU} measurements based on anchor availability, decayed \ac{UWB} quality, and motion context, under sparse, intermittent, and stale measurements.

\item We integrate a diffusion-based \ac{UWB} data augmentation strategy at the residual level to improve robustness and generalization under limited training data.

% \item We evaluate localization performance under varying anchor availability and demonstrate reliable trajectory estimation even with as few as two visible anchors, highlighting the suitability of the proposed method for highly constrained alpine and avalanche environments.

\item We validate the proposed approach using real-world experimental data and benchmark its performance against classical multilateration and loose-coupled fusion baselines using a high-precision \ac{GNSS} as ground truth.

\end{itemize}

The remaining structure of this paper is organized as follows:
\Cref{sec:related_work} reviews related work, including localization based on \ac{UWB} and \ac{IMU} and fusion algorithms comprising filter-based and learning-based fusion strategies.
\Cref{sec:system_overview} presents the system model and reviews state-of-the-art fusion approaches.
\Cref{sec:aoi-fusionnet} introduces our new \proposalName{} algorithm and provides details of the proposed model.
\Cref{sec:performance_evaluation} presents the evaluation results and discusses the lessons learned from our study.
Finally, \cref{sec:conclusion} summarizes the article and outlines future research directions.

\section{Related Work} \label{sec:related_work}

\subsection{UWB Range-based Localization}

Recent research has demonstrated that \Ac{UWB} has great potential for wireless ranging and localization, offering centimeter-level spatial resolution.
\ac{UWB} localization systems typically rely on \ac{ToF}, \ac{AoA}, \ac{TDoA} measurements with algorithms such as least square, Taylor algorithm, CHAN algorithm for position calculation.
The high positioning accuracy can be achieved under \ac{LOS} situations between the mobile node and the static anchors.
However, \ac{UWB} faces limitations in complex outdoor environments where \ac{NLOS} conditions deteriorates \ac{ToF}-based ranging.
Additionally, multipath effects and poor anchor placement (high \ac{GDOP}) further degrade positioning accuracy and hindering seamless localization performance \cite{feng2015gdop}.

\textcite{poulose2020accurate} proposed an \ac{UWB}-based positioning framework using \ac{EKF} to enhance indoor localization in a multi-anchor configuration.
The algorithm processes \ac{ToA} measurements from multiple anchors to address the inherent system nonlinearities, and its performance has been tested using MATLAB simulations in a 2D environment under both \ac{LOS} and \ac{NLOS} conditions.
% The results show that increasing the number of anchors improved localization accuracy, reduced computational complexity, and the results are compared against baseline techniques. 
While the proposed \ac{EKF}-based approach achieved positioning errors within a \SI{+-50}{\centi\meter}, its use in dynamic scenarios is constrained by its reliance on predefined noise models, thus limiting its adaptability in real-world conditions.

\textcite{van-herbruggen2024real-time} introduced two heuristic anchor selection algorithms for \ac{TWR}-based \ac{UWB} indoor positioning systems to improve accuracy and update rates.
One method selects anchors based on link quality and \ac{GDOP}, while the other method is designed as a lightweight, real-time algorithm for deployment on low-power \ac{UWB} tags.
The performance has been evaluated using a Kalman filter in industrial environments, achieving positioning accuracy in the order of \SIrange{15}{20}{\centi\meter}.
However, the non-constrained algorithm requires continuous communication with all anchors, and these methods depend on static link quality metrics without accounting for motion dynamics or temporal variations, which is essential for reliable localization in dynamic environments.

\subsection{IMU-based Localization}

Data-driven approaches using \ac{IMU} sensor have shown promising results for localization based on \ac{PDR} and have been investigated to improve individual components of \ac{PDR}.
\textcite{tiwari2023step} introduced a step detection method using the \ac{DWIFCM} algorithm for pedestrian dead-reckoning navigation, where it exploits features extracted from tri-axial accelerometer data and dynamically adjusts feature weights corresponding to changes in walking patterns and terrain.
\textcite{liu2023pedestrian} proposed an F-LSTM-based heading estimation technique that fuses \ac{IMU} sensor and magnetometer data to correct \ac{PDR} heading angle estimation.
This fully connected \ac{LSTM} model learns to regress the non-linear sensor error when it is trained on data collected in different heading directions.

While these algorithms demonstrate improved inertial navigation accuracy, the dependence on supervised training with restricted motion patterns and limited environmental variations, such as dynamic outdoor environments, limits their adaptability and generalization to unseen users and magnetic disturbances.
To address gyrometer saturation in orientation estimation, \textcite{neurauter2023motion} proposed two fusion algorithms using low-cost \ac{IMU} gyrometer and magnetometer readings, later fused with the state-of-the-art Madgwick's algorithm.
Despite testing on real snow avalanche datasets, the dependence on magnetometer data limits its robustness in environments sensitive to magnetic
disturbances.

\subsection{Model-based Sensor Fusion}

\ac{IMU} provides good localization performance over short step sequences, and several studies have explored its potential in multisensor fusion combining \ac{IMU} with other sensor modalities, including vision~\cite{beauvisage2022odometry}, LiDAR~\cite{zou2021comparative}, or UWB~\cite{guan2021fusing} to achieve better localization accuracy.
Among wireless communication technologies in outdoor scenarios, \ac{UWB} stands out as a cost-effective and compact solution capable of maintaining sub-meter accuracy.
Classical filtering methods such as \ac{KF} and its nonlinear variants have been extensively used so far \cite{li2024correntropy}.
Nonetheless, the reliance of such models on Gaussian noise models reduces their usefulness in nonlinear and non-Gaussian settings.
Hybrid solutions such as \cite{xu2025uvtrack,hu2024tightly}  that integrate vision with \ac{UWB} further improve robustness but are sensitive to avalanche-like environments where other factors, such as visibility and occlusions due to the surrounding powder cloud, adversely affect localization accuracy.

\textcite{xu2020seamless} introduced a loosely coupled integrated localization system based on a least square \ac{SVM} assisted \ac{UFIR}; however, the \ac{NLOS} situation is not considered.
\textcite{fan2025coupled} proposed an \ac{EKF} based classical tight coupling algorithm for indoor navigation that integrates raw \ac{UWB} with \ac{IMU} to mitigate \ac{NLOS} effects.
While the proposed EKF-based approach combines carrier motion characteristics to improve robustness, the manually orchestrated probabilistic models and threshold-dependent \ac{NLOS} handling limit its performance for outdoor scenarios. 
Similarly, \textcite{li2024tightly} introduced a tightly coupled indoor localization \ac{UWB}-\ac{PDR} solution using a factor-graph. 
The proposed method improves localization accuracy by jointly optimizing raw \ac{UWB} ranges and pedestrian motion constraints; however, it lacks adaptive noise modeling to account for environmental dynamics and its application is limited to 2-D scenarios.
The explicit assumption on motion dynamics and measurement noise statistics underlying classical model-based frameworks becomes difficult to satisfy under sparse anchor availability and multipath-induced non-Gaussian \ac{UWB} errors, thereby limiting their robustness in our scenario and motivating the data-based approach.

\subsection{Learning-based Sensor Fusion}

Recent studies suggest that data-driven learning approaches help to improve sensor fusion and localization performance compared to traditional model-based filtering, particularly in complex, non-linear, and non-Gaussian environments.

Machine learning techniques have been investigated to fuse measurements from multiple sensors and obtain precise localization.
Using \ac{XGBoost}-enhanced fusion framework, \textcite{tommingas2025estimation} proposed fusing \ac{UWB} and \ac{GNSS} to handle indoor-outdoor transitions by predicting sensor confidence levels.
However, the approach relies on \ac{GNSS} availability and hand-crafted features, which limit its scalability.
Deep neural network architectures such as \ac{CNN}-\ac{LSTM} have been proposed by~\cite{zhi2021performance}, taking advantage of \ac{CNN} and \ac{LSTM} architectures simultaneously to improve the performance of the integrated navigation system during GPS outages but the reliance on \ac{IMU} makes the system susceptible to sensor drift and synchronization challenges.

\textcite{muthineni2025deep} proposed a cascaded \ac{DL} architecture for \ac{UWB}-\ac{IMU} fusion in industrial environments.
While the approach outperforms \ac{EKF} baseline, the model is trained for a fixed number of inputs and lacks adaptivity to multipath-rich outdoor settings with varying intermittent \ac{LOS}/\ac{NLOS} conditions.
\textcite{fontaine2024transfer} presented a \ac{TL}-based automatic optimization solution to develop a high precision \ac{UWB} localization system.
Similarly, \textcite{yang2025nlos} proposed a tight coupled-\ac{BERT} algorithm for accurate indoor positioning. 
\textcite{tran2022gan-based} proposed \ac{GAN}-based data augmentation technique for enhancing the training dataset so that the model extracts multiple features of \ac{CIR} to reduce the localization error, caused by \ac{NLOS} conditions.
However, these methods depend upon the extraction of channel features from \ac{CIR} to explicitly perform \ac{LOS}/{NLOS} classification for error correction.

\begin{figure*}
    \centering
    \includegraphics[width=\textwidth]{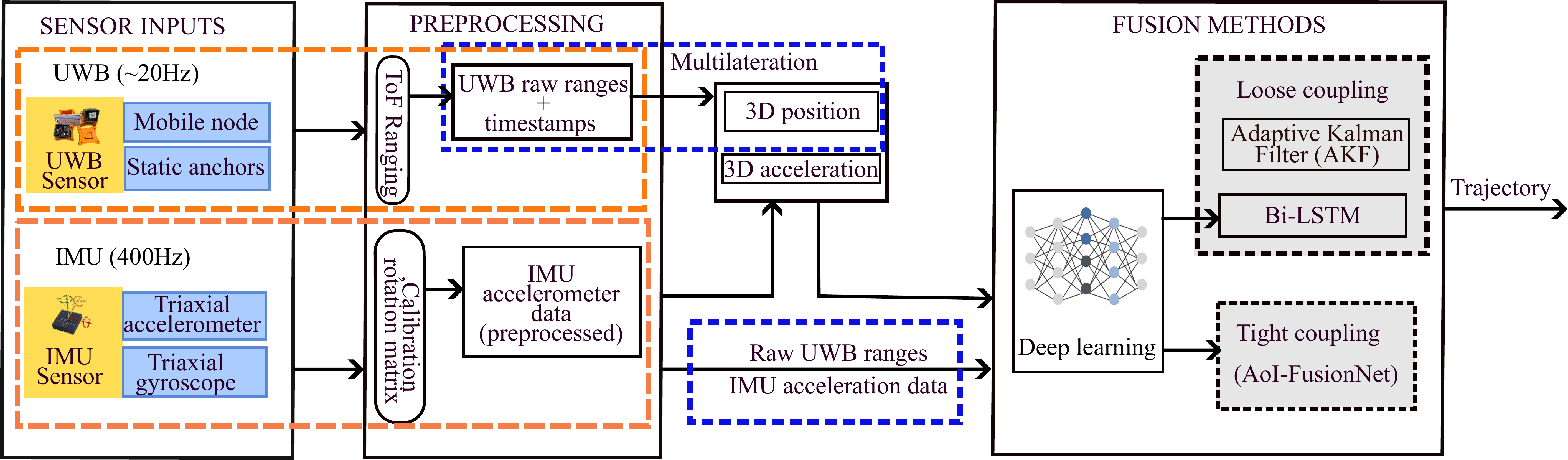}
    \caption{System model and data flow for UWB-IMU fusion.
    The mobile node is equipped with a UWB transceiver and a triaxial IMU and communicates with static anchors in a GNSS-denied outdoor environment.
    Raw UWB ranges with timestamps, and preprocessed inertial measurements are provided as inputs to multiple localization pipelines.
    Classical baselines include loose-coupled fusion using \ac{AKF} and a \ac{Bi-LSTM} model operating on trilaterated UWB positions.
    In contrast, the proposed {\proposalName} performs tightly coupled fusion directly on raw UWB and IMU measurements, incorporating AoI-aware decay, a learned fusion gate, and LSTM-based temporal modeling.
    The resulting 3D trajectory is evaluated against ground truth.}
    \label{fig:system-model}
    \vspace{-.8em}
\end{figure*}

\subsection{Our Solution: {\proposalName}}

In this work, we propose {\proposalName}, an \ac{AoI}-aware tightly coupled \ac{LSTM}-based \ac{UWB}-\ac{IMU} fusion framework designed for 3D trajectory estimation under intermittent rangings, sparse anchor visibility, high \ac{GDOP} and \ac{NLOS} conditions.
Unlike existing fusion architectures that rely solely on \ac{LOS} \ac{UWB} ranges by explicitly identify and exclude \ac{NLOS} measurements, our proposed architecture uses both \ac{LOS} and \ac{NLOS} measurements within a unified fusion process.
To the best of our knowledge, the explicit modeling of temporal measurement reliability within a tightly coupled learning-based fusion framework has not been investigated in this specific formulation.
The model integrates an \ac{AoI}-aware decay mechanism in the fusion algorithm that adaptively down-weighs stale \ac{UWB} measurements and a learned attention gate dynamically balances trust between \ac{UWB} and \ac{IMU} features based on the \ac{UWB} measurement freshness and anchor reliability.
% Our model learns temporal motion patterns directly from sequential sensor data instead of relying on \ac{GNSS} inputs, hand-crafted filters, or explicit \ac{NLOS} classification technique looking into \acp{CIR}. 
By modeling temporal dependencies through the recurrent architecture, our model learns temporal motion patterns directly from sequential sensor data and maintains robustness under sparse anchor measurements and \ac{NLOS} conditions. 
% The tight-coupled fusion technique outperforms \ac{UWB} trilateration-based position calculation, classical loose-coupled fusion of \ac{UWB} and \ac{IMU}, deep learning as well as traditional Bayesian filtering for outdoor avalanche-like environments. 

%
%
%

\section{System Model}\label{sec:system_overview}

In the following, we describe our system model that uses \ac{UWB} and \ac{IMU} systems as its key building modules. 
The fusion algorithm integrates \ac{UWB}-based ranging and inertial measurements to generate a multi-sensor dataset applicable for both classical filtering and \ac{DL}-based fusion methods.
As shown in \cref{fig:system-model}, the system adopts a modular architecture in which synchronized data collected from the mobile node and anchor network are followed by preprocessing and subsequently processed through sensor fusion and trajectory evaluation.
The system model is structured into four primary stages: (1) experimental setup mainly using mobile node (AvaNode) and static anchors (AvaAnchors) for data acquisition, (2) sensor-level preprocessing that includes \ac{UWB}-based ranging and localization, (3) \ac{IMU} calibration followed by frame transformation, (4) multi-sensor fusion methods involving loose-coupled \ac{UWB}-\ac{IMU} fusion, and tight-coupled \ac{UWB}-\ac{IMU} fusion.
Each stage of the system is described in detail in the following.

\subsection{AvaRange Scenario}

The sensor network deployed for this experiment comprises two principal components: the AvaNode \cite{neurauter2021evaluation}, which functions as the mobile sensing unit, and multiple AvaAnchors that are statically distributed across the test area to support position estimation \cite{kuss2024distributed}.
The sensor payload integrated into the AvaNode included an \ac{UWB} and an \ac{IMU}, and was further supported by a high-precision \ac{RTK} \ac{GNSS} receiver to provide reference data for ground-truth information.
In total, six anchors were deployed where one anchor was the gateway anchor responsible for managing the backhaul WiFi mesh network connecting the remaining anchors.
This configuration enables synchronized acquisition of multi-modal sensor data from \ac{UWB} module, an \ac{IMU}, and a high-precision RTK \ac{GNSS} sensor, for subsequent localization and sensor fusion analysis.

\begin{figure}
    \centering
    \includegraphics[width=0.8\columnwidth]{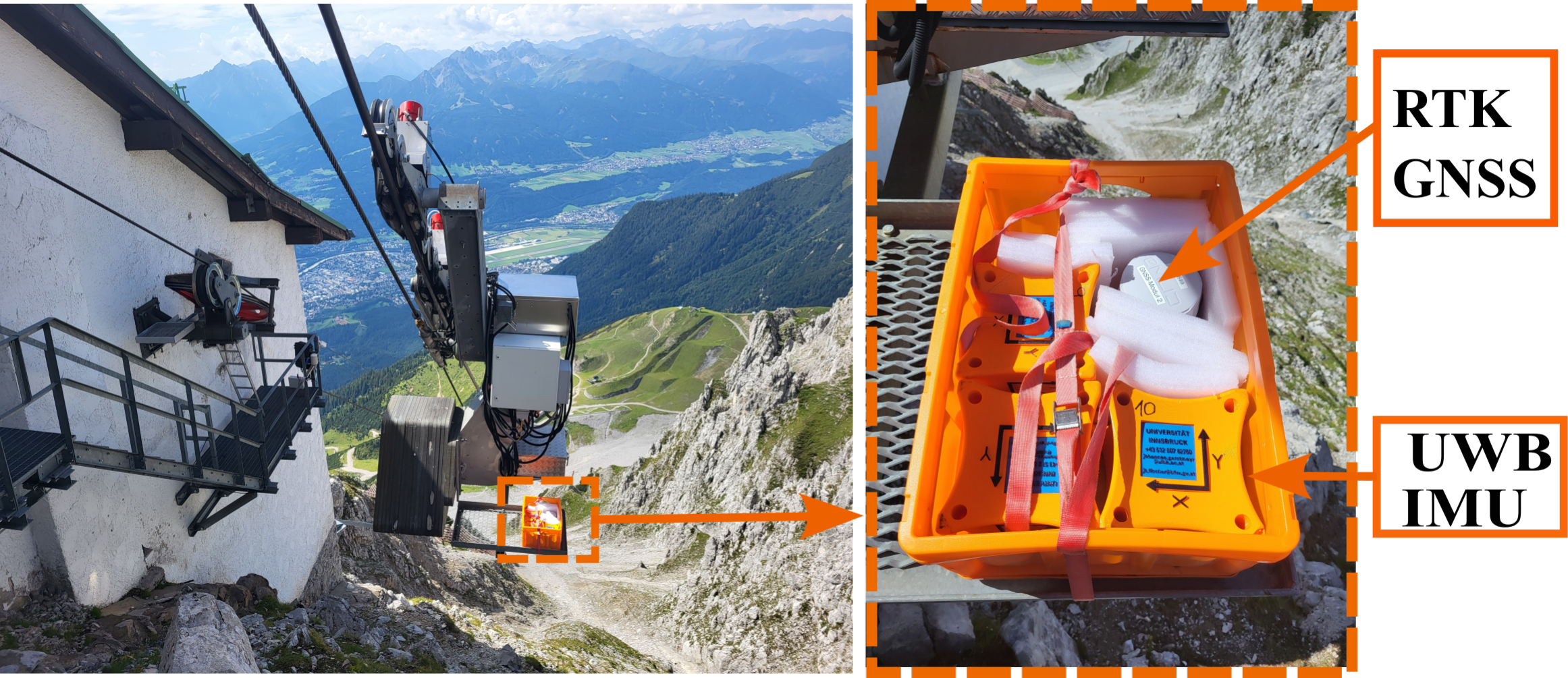}
    \caption{Tracking experiment at Innsbruck Nordkettenbahn cable car to derive the trajectory along the avalanche path. An IMU and UWB sensor node together with a reference GNSS is traversing in high alpine terrain that is equipped with static UWB anchors.}
    \label{fig:cabel_car}
    \vspace{-.8em}
\end{figure}

In this work, the measurement data for post-processing has been collected during a field experiment conducted in the Austrian Alps \cite{kuss2024distributed}.
% To evaluate the functionality of our system, a controlled outdoor validation experiment has been carried out near Innsbruck Nordkettenbahn.
The experimental arrangement consisted of an AvaNode mounted onto a cable car counterweight of the Hafelekar section of the Innsbruck Nordkettenbahn, which travels along the Nordkette avalanche test-site, and used herein to emulate realistic avalanche terrain geometry (see \cref{fig:cabel_car}). Moving nodes perform an experimental investigation of \ac{UWB}-based tracking under realistic alpine geometry.
The cable car provides a reproducible trajectory, thus allowing for comparison of measured trajectories against the ground truth.

\subsubsection{AvaNode}

The AvaNode represents the tracked device and is designed to simulate the motion of a snow granule within an avalanche.
It consists of multiple key sensing elements. \ac{UWB} \ac{ToF}-based ranging is performed using a Decawave DW1000 transceiver interfaced with an Adafruit Feather M0 microcontroller.
The DW1000 chip supports timestamping precision down to \SI{15}{\pico\second}, thus allowing highly accurate distance measurements.
Inertial sensing has been performed by MPU9250 \ac{IMU} that integrates tri-axial accelerometer and gyroscope.
The \ac{IMU} is housed within a 3D-printed cube with an edge length of \SI{160}{\milli\meter}, ensuring alignment between sensor coordinate frame and the geometric center of the enclosure.
\Ac{IMU} records motion data at a sampling frequency of \SI{400}{\hertz} and the measurements are logged to an onboard SD card via the Feather M0 microcontroller while, \ac{UWB} ranging data is sampled at a frequency of \SI{20}{\hertz}.
Despite the incorporation of additional components in the AvaNode platform that includes low-cost GPS module and a recovery mechanism, only the measurements collected from \ac{UWB} and \ac{IMU} sensors are focused in this article for tracking analysis.

\subsubsection{AvaAnchor}

Six AvaAnchors equipped with DW1000 \ac{UWB} chips serve as static anchors with known fixed locations installed along the cable car route.
In addition, the anchors establish an IEEE 802.11 WiFi mesh network while LoRa is used for sending commands to the node.
The anchor positions were precisely surveyed with \ac{RTK} \ac{GNSS} system to allow accurate multilateration.
In a nutshell, this configuration enables reliable acquisition of high-resolution motion data under controlled conditions to evaluate sensor fusion techniques for 3D trajectory reconstruction in avalanche-like environments.

\subsubsection{Ground truth}
% \textcolor{red}{RTK GNSS (Reach RS2+ from Emlid). spatial and temporal which resolution/accuracy? describe the cable car trajectory (speed, scale, no rotation, linear)}
To establish high-accuracy ground truth, an \ac{RTK} \ac{GNSS} (Emlid Reach RS2+) unit has been mounted.
% The \ac{RTK} \ac{GNSS} receiver provided the ground truth positions.
The reference positions were recorded at \SI{10}{\hertz}.
Over the synchronized segment, the start-to-end displacement is approximately \SIrange{700}{720}{\meter} (horizontal: $\approx 620\,\mathrm{m}$; vertical: $\approx 343\,\mathrm{m}$), corresponding to an average speed of $\approx 22\,\mathrm{km/h}$.

\subsection{UWB Ranging and Localization}

The \ac{AltDS-TWR} method \cite{neirynck2016alternative} is employed for \ac{UWB}-based distance estimation.
% This method improves ranging accuracy by exchanging three messages between the mobile node and anchor as shown in \cref{fig:alt_ds_twr}. 
% This method does not assume symmetric delays and remains robust in asynchronous systems, making it well-suited for practical deployments \cite{neirynck2016alternative}.
Even though \ac{AltDS-TWR} reduces timing uncertainty and mitigates clock drift without depending upon tight clock synchronization, it does not eliminate the multipath and NLOS bias.
The time-of-flight (\(T_f\)) is calculated as:
\begin{equation}
T_f = \frac{R_a R_b - D_a D_b}{2(R_a + D_a)},
\end{equation}
where \(R_a\) and \(R_b\) denote round-trip delays measured at each device, and \(D_a\), \(D_b\) are the respective known reply delays \cite{kuss2023measurement}.
Distances $d_i$ can be calculated from time-of-flight (\(T_f\)) measurements as:
\begin{equation}
d_i = c \cdot T_f,
\end{equation}
Here, \( c \) denotes the speed of light, which generally depends on temperature, humidity and the snow-air density inside a powder cloud, but here kept constant. 
% \textcolor{red}{$\tilde{d}$ from eq. [23] is missing here!}
This formulation results in high-accuracy rangings even in the presence of timing asymmetries.

% \begin{figure}
%     \centering
%     \includegraphics[width=0.9\columnwidth]{figs/ToF.pdf}
%     \caption{AltDS-TWR message exchange for time-of-flight estimation using three packets.}
%     \label{fig:alt_ds_twr}
%     \vspace{-.8em}
% \end{figure}

% The position of the mobile node from \ac{ToF} rangings is estimated through multilateration. Generally, trilateration computes the position of a target based on its measured \ac{UWB} ranges to at least three known anchors.
% In 2D, each distance describes a circle around an anchor.
% The intersection of three circles provides the target's location:
% \begin{equation}
%     d = \sqrt{(x_t - x_i)^2 + (y_t - y_i)^2}, \quad i = 1, 2, ..., N
% \end{equation}
% where \((x_i, y_i)\) are the known positions of anchors and \(d_i\) is the measured distance.

% In 3D, as shown in \cref{fig:3d_trilateration}, distances define spheres, and the position \((x_t, y_t, h_t)\) of the node is found by solving for the intersection of at least four spheres:
% \begin{equation}
%     d_i = \sqrt{(x_t - x_i)^2 + (y_t - y_i)^2 + (h_t - h_i)^2}, \quad i = 1, 2, ..., N
% \end{equation}
% where \((x_i, y_i, h_i)\) are the known positions of anchors and \(d_i\) is the measured distance.

% \begin{figure}
%     \centering
%     \includegraphics[width=0.8\columnwidth]{figs/trilateration 3D.drawio.pdf}
%     \caption{3D trilateration using spheres from multiple anchors to localize the mobile node in \((x, y, h)\).}
%     \label{fig:3d_trilateration}
%     \vspace{-.8em}
% \end{figure}

For the calculation of the AvaNode position, we used multilateration formulated as a non-linear optimization problem.
Specifically, the objective of this optimization is to minimize the \ac{MSE} between the measured \ac{UWB} distances and the corresponding Euclidean distances to a set of fixed AvaAnchors.
Let $\mathbf{a}_i = (x_i, y_i, z_i)$ denote the known coordinates of anchor $i$, and let $d_i$ represent the corresponding measured distance.
The estimated position of the mobile node $\mathbf{p} = (x, y, z)$ is obtained by solving the following problem:
\begin{equation}
\text{MSE}(\mathbf{p}) = \frac{1}{N} \sum_{i=1}^{N} \left( \| \mathbf{p} - \mathbf{a}_i \| - d_i \right)^2
\end{equation}
where $N$ represents the number of anchors providing distance measurements. 
A minimum of four anchors is required to obtain a unique solution in three-dimensional space.

\begin{figure}
    \centering
    \includegraphics[width=.8\columnwidth]{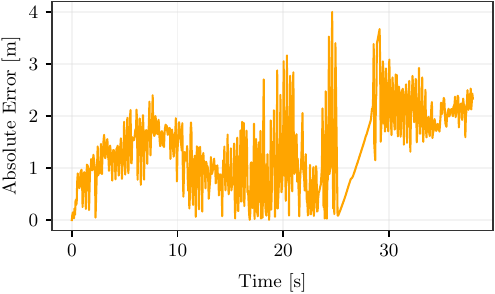}
    \vspace{-.8em}
    \caption{Relative error computed between UWB-only trilaterated positions and RTK GNSS ground truth along Z-axis.}
    \label{fig:uwb_vs_rtk_all}
    \vspace{-.8em}
\end{figure}

\Cref{fig:uwb_vs_rtk_all} shows the absolute localization error along the vertical axis for \ac{UWB}-only multilateration against the \ac{RTK} \ac{GNSS} reference.
Over time, an increasing deviation of up to \SI{4}{\meter} is observed, particularly along the vertical axis, where the impact of unfavorable anchor geometry reflected by high \ac{GDOP} is more significant.

\subsection{IMU Calibration and Preprocessing}

The calibration of \ac{IMU} has been adopted from the procedure proposed by \textcite{neurauter2022novel,winkler2024particle}.
These methods explicitly model sensor errors, estimate orientation from angular velocity measurements, and compute gravity-compensated integration of accelerometer measurements.
Precise calibration of \ac{IMU} sensors is imperative for mitigation of deterministic measurement errors.
After proper calibration, the accelerometer data reflects only inertial and gravitational components, while the gyroscope accurately tracks angular motion independent of any systematic drift. 

\begin{figure}
    \centering
    \includegraphics[width=.8\columnwidth]{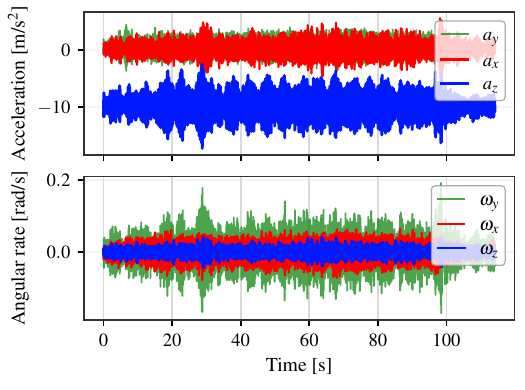}
    \vspace{-.8em}
    \caption{Calibrated accelerometer and gyroscope measurements along X-, Y-, and Z-axes.}
    \label{fig:calibrated}
    \vspace{-.8em}
\end{figure}

The resulting calibrated inertial measurements are illustrated in \cref{fig:calibrated}. 
During the experiment, the \ac{IMU} sensor experienced predominantly translational motion with limited rotational activity.
Therefore, the accelerometer data exhibits consistent behavior along X- and Y- axes, however, the measurements along Z-axis are largely influenced by gravity and centered around \SI{-9.81}{\meter\per\second\squared}, consistent with theoretical expectations.
In contrast, the gyroscope measurements remain close to zero across all axes, with only minor fluctuations, highlighting the absence of significant rotational motion.

Following calibration, the next stage is motion reconstruction, where inertial data, specifically translational acceleration and angular velocity, are used to reconstruct the 3D trajectory of the mobile node.
The complete kinematic state of the mobile node is estimated by integrating calibrated accelerometer and gyroscope measurements over time.
To transform measurements from the local sensor (body) frame to the global frame of reference, rotation matrices are computed that allows orientation estimation from gyroscope data.
Based on the estimated orientation and corresponding rotation matrices, the linear motion is reconstructed by the transformation of calibrated translational acceleration to the global coordinate frame of reference. 
Afterwards, gravitational acceleration is subtracted to obtain translational acceleration, which is followed by the numerical integration of global acceleration using the explicit Euler method to obtain velocity and position.
% The resulting gravity-compensated translational acceleration and motion estimates form a globally consistent 3D trajectory, which can be directly compared with external localization references such as \ac{UWB} or \ac{GNSS} systems.
The resulting gravity-compensated translational acceleration is used as an input to fusion methods.

\subsection{Loose-coupled UWB-IMU Fusion Using Kalman Filter}

We first adopted an \ac{AKF}, an extension of standard \ac{KF}, as our baseline method for loose coupling-based fusion of data from \ac{UWB} and \ac{IMU} sensors.
By dynamically adjusting the measurement noise covariance under time-varying sensor noise and uncertainties, the filter estimates 3D position, velocity, and acceleration.
The state vector can be defined as:
\begin{equation}
\mathbf{x}_k =
\begin{bmatrix}
x_k & y_k & z_k & \dot{x}_k & \dot{y}_k & \dot{z}_k & \ddot{x}_k & \ddot{y}_k & \ddot{z}_k
\end{bmatrix}^\top
\end{equation}
State prediction follows the standard \ac{KF} prediction model as:
\begin{align}
\mathbf{x}_{k|k-1} &= \mathbf{F} \mathbf{x}_{k-1|k-1}, \\
\mathbf{P}_{k|k-1} &= \mathbf{F} \mathbf{P}_{k-1|k-1} \mathbf{F}^\top + \mathbf{Q}_k,
\end{align}
where \(\mathbf{F}\) is the constant acceleration \(9 \times 9\) state transition matrix based on sampling interval \(\Delta t\), and \(\mathbf{Q}_k = \mathbf{G} \mathbf{Q}_0 \mathbf{G}^\top\) is the process noise covariance.
Measurement updates incorporating actual sensor data are performed using:
\begin{equation}
\mathbf{z}_k = \mathbf{H}_k \mathbf{x}_{k|k-1} + \mathbf{v}_k, \quad \mathbf{v}_k \sim \mathcal{N}(0, \mathbf{R}_k).
\end{equation}
The Kalman gain and state update equations are following:
\begin{align}
\mathbf{y}_k &= \mathbf{z}_k - \mathbf{H}_k \mathbf{x}_{k|k-1}, \\
\mathbf{S}_k &= \mathbf{H}_k \mathbf{P}_{k|k-1} \mathbf{H}_k^\top + \mathbf{R}_k, \\
\mathbf{K}_k &= \mathbf{P}_{k|k-1} \mathbf{H}_k^\top \mathbf{S}_k^{-1}, \\
\mathbf{x}_{k|k} &= \mathbf{x}_{k|k-1} + \mathbf{K}_k \mathbf{y}_k, \\
\mathbf{P}_{k|k} &= (\mathbf{I} - \mathbf{K}_k \mathbf{H}_k) \mathbf{P}_{k|k-1}.
\end{align}
To account for time-varying uncertainties, \ac{AKF} adaptively updates measurement noise covariance \(\mathbf{R}_k\) using the empirical innovation covariance over a sliding window of size \(N\).
Let $\mathbf{y}_i$ be the innovation vector at time step $i$ over a fixed window of size $N$.
The sample innovation covariance is:
\begin{equation}
\hat{\mathbf{C}} = \frac{1}{N - 1} \sum_{i=k-N+1}^{k} \mathbf{y}_i \mathbf{y}_i^\top
\end{equation}
We update the measurement noise covariance as:
\begin{equation}
\mathbf{R}_k = \hat{\mathbf{C}} - \mathbf{H}_k \mathbf{P}_{k|k-1} \mathbf{H}_k^\top
\end{equation}
In addition, \ac{GDOP}-based dynamic scaling to $\mathbf{R}_k$ is applied to \ac{UWB} measurement noise to account for geometric uncertainty.
\begin{equation}
\mathbf{R}_{\text{UWB}}^{\text{scaled}} = \alpha(g)^2 \cdot \mathbf{R}_{\text{UWB}}
\end{equation}
where the scaling factor $\alpha(g)$ is a function of the \ac{GDOP} value $g$.
This combined residual-based adaptation and geometry-aware scaling improves reliability under degraded anchor configurations, especially along the vertical axis, where \ac{UWB} measurements are often less accurate, due to the limited height variation in anchor positions.

\subsection{Loose-coupled UWB-IMU Fusion using Bi-LSTM} % (LC(Bi-LSTM))}

In our preliminary work~\cite{bibi2025deep}, we proposed a learning-based loose-coupling fusion technique using \ac{Bi-LSTM}-based deep recurrent network on the very same experimental dataset.
In this method, \ac{UWB} position estimates are fused with \ac{IMU}-derived inertial measurements, consistent with loose-coupling pipeline. 
The designed model learns the complex temporal relationships between \ac{IMU} acceleration measurements and position estimates obtained from \ac{UWB} multilateration.
The model thereby learns to predict the frame-wise displacement vectors referred to as the delta position $\Delta \mathbf{p}_t$ between consecutive timestamps, where each frame contains a single-time instance incorporating input features that contain both \ac{IMU} and \ac{UWB} data.
Instead of predicting absolute positions directly, the model focuses on learning incremental motion patterns to improve generalization and reduces long-term drifts. 

Input sequences are taken in the sliding window of $200$ consecutive time steps, where each time step contains six features that are three-axis accelerometer readings and also three-axis \ac{UWB} positions, creating a $[200 \times 6]$ input tensor.
The architecture consists of three stacked \ac{Bi-LSTM} layers, each layer has a hidden dimension of 512 units per direction (forward and backward), followed by the fully connected layer producing 3D displacement vectors.
During inference, these predicted displacement deltas are cumulatively summed to reconstruct a full 3D trajectory.
The training of the model is performed using \ac{Adam} optimizer with a learning rate of $0.001$ and mini-batch size of $64$.
Early stopping with a patience of $30$ epochs prevents model from overfitting.
To account for axis-dependent error characteristics, a \ac{WMSE} loss function is applied, placing greater emphasis on the vertical dimension.
It can be expressed as follows:
% \begin{equation}
% \label{eq:wmse}
% loss(x, y, z) = w_x (\hat{x} - x)^2 + w_y (\hat{y} - y)^2 + w_z (\hat{z} - z)^2
% \end{equation}
% where, \( \hat{x}, \hat{y}, \hat{z} \) denotes the predicted displacement components, and \( x, y, z \) are the corresponding ground truth values.
% The weighting coefficients \( w_x, w_y, w_z \) are selected to reflect axis-dependent importance during training.
\begin{equation}
\label{eq:wmse}
\mathcal{L}_{\mathrm{WMSE}}(t)
=
\bigl\|
\mathbf{W}^{1/2}
\bigl(
\widehat{\Delta\mathbf{p}}_t - \Delta\mathbf{p}_t
\bigr)
\bigr\|_2^2 ,
\end{equation}
where $\widehat{\Delta\mathbf{p}}_t=[\widehat{\Delta x}_t,\widehat{\Delta y}_t,\widehat{\Delta z}_t]^T$ and
$\Delta\mathbf{p}_t=[\Delta x_t,\Delta y_t,\Delta z_t]^T$ denotes the predicted and ground-truth incremental displacement vectors, respectively, and $\mathbf{W}=\mathrm{diag}(w_x,w_y,w_z)$ shows a diagonal weighting matrix that emphasizes axis-dependent errors.
For a mini batch of size \( N \), the averaged \ac{WMSE} $\mathcal{L}$ can be calculated as:
% \begin{equation}
% \label{eq:batchwmse}
% \mathcal{L} = \frac{1}{N} \sum_{n=1}^{N} \left[ w_x (\hat{x}_n - x_n)^2 + w_y (\hat{y}_n - y_n)^2 + w_z (\hat{z}_n - z_n)^2 \right]
% \end{equation}
\begin{equation}
\label{eq:batchwmse}
    \mathcal{L}=\frac{1}{N}\sum_{n=1}^{N}\mathcal{L}_{\mathrm{WMSE}}(\widehat{\Delta\mathbf{p}}^{(n)},\Delta\mathbf{p}^{(n)}).
\end{equation}

% The process is summarized in \cref{alg1:lc(bi-lstm)}.
% The training hyperparameters are presented in \cref{tab:Bi-LSTM-hyperparams}.
The model configuration was empirically optimized to find the sweet spot and achieve optimal performance.
More details can be found in our earlier work \cite{bibi2025deep}.

\section{{\proposalName}}\label{sec:aoi-fusionnet}

Contrary to loose-coupling (LC-Bi-LSTM), a tightly coupled learning-based localization framework ({\proposalName}) is proposed that combines inertial and \ac{UWB} ranging data while explicitly accounting for ranging sparsity, temporal staleness, and availability of anchors.
This method is specifically designed to operate under \ac{UWB} degradation, including scenarios where only one or two anchors are intermittently visible.
{\proposalName} operates on the sliding window of multi-sensor temporal measurements and performs incremental motion prediction rather than absolute position estimation.
Such an approach mitigates long-horizon drift and improves localization performance under missing or unreliable \ac{UWB} measurements.
The predicted motion deltas are then accumulated to recover the full trajectory.

Before feeding inertial data into the tightly coupled fusion pipeline, the acceleration bias optimization method proposed by \textcite{neurauter2022novel} is applied to mitigate the integration drift in the calibrated translational acceleration data.
The measured acceleration is corrected by first-order polynomial in time:
\[
\tilde{\mathbf{a}}(t) = \mathbf{a}(t) + \mathbf{a}_0 + \mathbf{a}_1 t,
\]
where $\mathbf{a}_0, \mathbf{a}_1 \in \mathbb{R}^3$ are the optimized correction parameters. 
Rotation matrices are then applied to the corrected acceleration so that it is rotated into the global frame of reference, after which global velocity and position are obtained via numerical integration.
The correction parameters $\mathbf{a}_0$ and $\mathbf{a}_1$  are selected by minimizing a cost function:
\[
F = \|\mathbf{v}_{\text{final}}\|^2 + \|\mathbf{x}_{\text{final}} - \mathbf{x}_{\text{ref}}\|^2,
\]
where $\mathbf{v}_{\text{final}}$ and $\mathbf{x}_{\text{final}}$ denote the terminal velocity and position, while $\mathbf{x}_{\text{ref}}$ is a known reference endpoint.
This optimization encourages that the reconstructed motion ends at rest and matches with the known final position, thereby removing the sensor drift.

\subsection{\proposalName{} Model Construction}

\begin{table}
    \centering
    \caption{Key Hyperparameters of {\proposalName}: Tightly-Coupled IMU-UWB Fusion Model}
    \label{tab:aoi_fusionnet_hyperparams}
    \begin{tabular}{lll}
        \toprule
        \textbf{Parameter} & \textbf{Value} & \textbf{Description} \\
        \midrule
        IMU input dim. & 3 & 3-axis accel. \\
        UWB anchors ($A$) & Variable & Available anchors \\
        Output dim. & 3 & $\Delta x,\Delta y,\Delta z$ \\
        Hidden size ($H$) & 128 & Shared latent dim. \\
        Temporal model & LSTM & Tight temporal coupling \\
        Number of Layers & 1 & LSTM layer \\
        Anchor embed. ($E$) & 3 & Anchor identity encoding \\
        Learned attention  & Yes & IMU-UWB gating \\
        Loss function & Composite Huber & Robust multi-term loss \\
        Optimizer & AdamW & Weight-decayed Adam \\
        Learning rate & $3\times10^{-4}$ & Initial LR \\
        Early stopping & 20 epochs & Val-based patience \\
        Data split & 70/15/15 & Train/Val/Test \\
        \bottomrule
    \end{tabular}
    \vspace{-.8em}
\end{table}

{\proposalName} follows a four-stage processing pipeline designed to (1) encode modality-specific features,
(2) model the time-varying reliability of \ac{UWB} measurements using \ac{AoI},
(3) adaptive cross-modal fusion in a reliability-aware manner over time,
and (4) predict incremental motion updates.
This fusion framework includes an explicit reliability model for \ac{UWB} measurements using learned \ac{AoI}-decay, reducing impact of stale ranging updates.
The age-weighted \ac{UWB} features are integrated with \ac{IMU} data through a learned attention gate that modulates the relative contribution of each sensor modality based on measurement availability, temporal freshness, and motion context.
The fused sequence is then processed by a recurrent model to enforce temporal dependencies, while anchor-agnostic generalization is achieved via learned anchor embeddings, allowing the framework to adapt to varying anchor configurations without retraining.

% Finally, anchor-agnostic generalization is achieved via learned anchor embeddings, allowing the framework to adapt to varying anchor configurations without retraining.

\Cref{alg2:aoi_fusionnet_full} presents the pseudo code for the proposed model and \cref{tab:aoi_fusionnet_hyperparams} summarizes its hyperparameters.

\begin{algorithm}[t]
\caption{Pseudo code for tightly-coupled \ac{UWB}-\ac{IMU} Fusion: {\proposalName}}
\label{alg2:aoi_fusionnet_full}
\begin{algorithmic}[2]
\Require IMU $\{\mathbf{u}_t\}_{t=1}^{T}$, UWB ranges $\{d_{t,a}\}_{t=1,a=1}^{T,A}$, mask $\{m_{t,a}\}_{t=1,a=1}^{T,A}$, AoI $\{\tau^{\mathrm{age}}_{t,a}\}_{t=1,a=1}^{T,A}$, window length $T$
\Require (Training) ground-truth trajectory $\mathbf{P}=\{\mathbf{p}_t\}_{t=1}^{T+1}$ from RTK-GNSS, optimizer $\mathrm{Opt}$, scheduler $\mathrm{Sch}$
\Ensure (Training) best parameters $\theta^\star$;\quad (Testing) reconstructed positions $\{\hat{\mathbf{p}}_t\}_{t=1}^{T+1}$

\State \textbf{Training Phase}
\State Split data into train/val/test; compute train-only normalization $(\mu,\sigma)$
\State Initialize fusion network $f_\theta$ (\ac{LSTM}) with parameters $\theta$

\For{epoch $e=1$ to $E$}
    \State Set model to train mode
    \For{each mini-batch of windows}
        \State Extract $(\mathbf{U},\mathbf{D},\mathbf{M},\boldsymbol{\tau},\mathbf{P})$
        \State Causal-fill and normalize $\mathbf{D}$
        \State $\Delta\hat{\mathbf{P}} \gets f_\theta(\mathbf{U},\mathbf{D},\mathbf{M},\boldsymbol{\tau})$
        \State $\hat{\mathbf{p}}_1 \gets \mathbf{p}_1$
        \For{$t=1$ to $T$}
            \State $\hat{\mathbf{p}}_{t+1} \gets \hat{\mathbf{p}}_{t} + \Delta\hat{\mathbf{p}}_{t}$
        \EndFor
	 \State Compute total loss $\mathcal{L}$ and update $\theta$ via $\mathrm{Opt}$
    \EndFor
    \State Validate endpoint error; save best $\theta^\star$
    \State Update learning rate via $\mathrm{Sch}$
\EndFor

\State \textbf{Testing Phase}
\State Load $\theta^\star$ and set eval mode
\For{each test window}
    \State Extract $(\mathbf{U},\mathbf{D},\mathbf{M},\boldsymbol{\tau})$
    \State Causal-fill and normalize $\mathbf{D}$
    \State $\Delta\hat{\mathbf{P}} \gets f_{\theta^\star}(\mathbf{U},\mathbf{D},\mathbf{M},\boldsymbol{\tau})$
    \State Set $\hat{\mathbf{p}}_1 \gets \mathbf{0}$ \Comment{relative trajectory within the window}
    % \State Integrate $\Delta\hat{\mathbf{P}}$ to obtain $\{\hat{\mathbf{p}}_t\}$
    \For{$t=1$ to $T$}
        \State $\hat{\mathbf{p}}_{t+1} \gets \hat{\mathbf{p}}_{t} + \Delta\hat{\mathbf{p}}_{t}$
    \EndFor
\EndFor
\State \Return reconstructed trajectory $\{\hat{\mathbf{p}}_t\}_{t=1}^{T+1}$
\end{algorithmic}
\end{algorithm}

\subsubsection{IMU Feature Encoding}
The proposed model operates on synchronized input data from \ac{UWB} and \ac{IMU} sensors over a temporal window of length $T$.
At each discrete time step $t$, the model processes optimized tri-axial acceleration measurements $\mathbf{u}_t\ \in \mathbb{R}^{3}$.
% where $D_{\mathrm{imu}}$ represents the dimensionality of the \ac{IMU} feature vector that includes translational acceleration.
% The complete representation of \ac{IMU} sequence over the window can be denoted by \todo{something is broken here}
% $\mathbf{U} = \{\mathbf{u}_t\}_{t=1}^T$.
% \begin{equation}
% \mathbf{h}^{\mathrm{imu}}_t = f_{\mathrm{imu}}(\mathbf{u}_t),
% \qquad
% \mathbf{h}^{\mathrm{imu}}_t \in \mathbb{R}^{H},
% \end{equation}
% where $f_{\mathrm{imu}}(\cdot)$ refers to a shared \ac{IMU} encoder implemented using a multilayer perceptron.
The complete representation of \ac{IMU} sequence over the window can be denoted by
\begin{equation}
\mathbf{U} = \{\mathbf{u}_t\}_{t=1}^{T}.
\end{equation}

A shared IMU encoder $f_{\mathrm{imu}}(\cdot)$ is applied independently at each time step to obtain latent features $\mathbf{h}^{\mathrm{imu}}_t \in \mathbb{R}^{H}$ as
\begin{equation}
\mathbf{h}^{\mathrm{imu}}_t = f_{\mathrm{imu}}(\mathbf{u}_t),
\end{equation}
where $f_{\mathrm{imu}}(\cdot)$ is implemented as a multilayer perceptron with parameters shared across time and $\mathbf{h}^{\mathrm{imu}}_t$ is the learned feature embedding of IMU acceleration data used for fusion.
The encoded sequence is therefore
\begin{equation}
\mathbf{H}^{\mathrm{imu}} = \{\mathbf{h}^{\mathrm{imu}}_t\}_{t=1}^{T}.
\end{equation}

\subsubsection{UWB Representation with \ac{AoI}-Based Reliability}
In parallel, the network receives \ac{UWB} ranging measurements to a set of fixed anchors indexed by $\mathcal{A} = \{1,\dots,N_A\}$ where $N_A = |\mathcal{A}|$ is the number of deployed anchors.
These measurements are arranged into a matrix
\begin{equation}
    \mathbf{D} = \{d_{t,a}\} \in \mathbb{R}^{T \times N_A}, \quad a \in \mathcal{A}.
\end{equation}
where $d_{t,a}$ is the measured distance between the mobile node and anchor $a$ at time step $t$.
However, it is possible that \ac{UWB} measurements may be missing or unreliable due to intermittent communication and \ac{NLOS} conditions.
To explicitly model \ac{UWB} availability, a binary measurement mask
\begin{equation}
    \mathbf{M} = \{m_{t,a}\} \in \{0,1\}^{T \times N_A}, \quad a \in \mathcal{A}.
\end{equation}
is introduced, where $m_{t,a}=1$ indicates the presence of valid \ac{UWB} measurement and $m_{t,a}=0$ denotes a missing or invalid measurement.
This mask will later be combined with \ac{AoI} decay to estimate the time-varying reliability of \ac{UWB} input.

The proposed model also accounts for the temporal freshness of \ac{UWB} rangings using \ac{AoI} representation.
In this context, \ac{AoI}  measures the elapsed time since the last available \ac{UWB} range update from each anchor.
For each anchor $a \in \mathcal{A}$ and time step $t$, the \ac{AoI} $\tau^{\mathrm{age}}_{t,a}$  can be defined as
\begin{equation}
    \boldsymbol{\tau} = \{ \tau^{\mathrm{age}}_{t,a} \} \in \mathbb{R}_{+}^{T \times N_A},
\end{equation}
where $\mathbb{R}_{+}$ represents the set of non-negative real numbers. 
The \ac{AoI} variable is defined recursively for each
anchor $a \in \mathcal{A}$. As, $m_{t,a} \in \{0,1\}$ denote the \ac{UWB} measurement availability indicator. The AoI evolves as
\begin{equation}
\tau^{\text{age}}_{t,a} =
\begin{cases}
0, & \text{if } m_{t,a}=1, \\
\tau^{\text{age}}_{t-1,a} + 1, & \text{if } m_{t,a}=0,
\end{cases}
\end{equation}
with initialization $\tau^{\text{age}}_{0,a}=0$.
The resulting AoI tensor $\boldsymbol{\tau} = \{\tau^{\text{age}}_{t,a}\}$
provides an explicit, physically interpretable measure of temporal freshness, enabling the network to attenuate stale \ac{UWB} observations during fusion.

% While preserving anchor-specific measurement characteristics and enabling anchor-agnostic generalization, {\proposalName} assigns 
Each anchor $a \in \mathcal{A}$ is assigned a learnable embedding vector $\mathbf{e}_a \in \mathbb{R}^{E}$, where $E$ denotes the dimensionality of the anchor embedding space and it is set to $E=3$, as this provided stable performance without gains from larger embeddings.
These anchor embeddings allow the network to capture systematic anchor-dependent biases rather than explicitly encoding anchor coordinates or identities.
To handle intermittent measurements, missing values in $d_{t,a}$ are causally filled by propagating the last valid observation, producing $\bar{d}_{t,a}$.
The resulting signal is standardized using per-anchor statistics computed from the training data:
\begin{equation}
\tilde{d}_{t,a} =
\frac{\bar{d}_{t,a} - \mu_a}{\sigma_a + \varepsilon},
\end{equation}
where $\mu_a$ and $\sigma_a$ are the training-set mean and standard deviation for anchor $a$, and $\varepsilon$ is a small constant for numerical stability.

The composite \ac{UWB} feature is then formed as
\begin{equation}
\mathbf{x}^{\mathrm{uwb}}_{t,a} =
\Big[
\tilde{d}_{t,a},\;
\exp\!\left(-\tau^{\mathrm{age}}_{t,a} / \lambda_a\right),\;
\mathbf{e}_a
\Big],
\end{equation}
where $\tau^{\mathrm{age}}_{t,a}$ is the observed age of the most recent measurement and $\lambda_a$ is a learned anchor-specific decay constant controlling the influence of stale observations.
% For each time step $t$ and anchor $a$, a composite \ac{UWB} feature vector is constructed as
% \begin{equation}
%     \mathbf{x}^{\mathrm{uwb}}_{t,a} =
%     \Big[
%     \tilde{d}_{t,a},\;
%     \exp\!\left(-\tau^{\mathrm{age}}_{t,a} / \tau_a  \right),\;
%     \mathbf{e}_a
%     \Big],
% \end{equation}
% where $\tilde{d}_{t,a}$ denotes the normalized \ac{UWB} range measurement.
% The normalized range $\tilde{d}_{t,a}$, which is used as input to the network, is obtained via per-anchor z-score standardization:
% \begin{equation}
% \tilde{d}_{t,a} =
% \frac{\bar{d}_{t,a} - \mu_a}{\sigma_a + \varepsilon},
% \end{equation}
% where $\mu_a$ and $\sigma_a$ are the mean and standard deviation computed from the training-set measurements of anchor $a$, and $\varepsilon$ is a small constant for numerical stability.
% $\tau^{\mathrm{age}}_{t,a}$ refers to the age of the most recent measurement from anchor $a$, and $\tau_a$ is a learned anchor-specific decay constant that controls the attenuation of stale \ac{UWB} observations. 
% Note that $\tau^{\mathrm{age}}_{t,a}$ is the observed age, while $\tau_a$ is a learnable decay timescale specific to anchor $a$. 

Then, a shared anchor-wise encoder $f_{\mathrm{uwb}}(\cdot)$ is applied to each feature vector, producing a latent representation $\mathbf{h}^{\mathrm{uwb}}_{t,a} \in \mathbb{R}^{H}$ as
\begin{equation}
    \mathbf{h}^{\mathrm{uwb}}_{t,a} = f_{\mathrm{uwb}}(\mathbf{x}^{\mathrm{uwb}}_{t,a}),
\end{equation}
where $H$ denotes the size of the learned latent representation used by the network internally.
To account for measurement availability and temporal freshness, an \ac{AoI}-weighted reliability coefficient is defined as
\begin{equation}
\tilde{m}_{t,a}
=
m_{t,a}
\exp\!\left(
-\frac{\tau^{\mathrm{age}}_{t,a}}{\lambda_a}
\right),
\end{equation}
which models reliability decay as the measurement age increases. This exponential form is consistent with the forgetting behavior used in recursive state estimation, providing continuous attenuation of stale observations while preserving recent measurements.
These anchor-level features are then aggregated across anchors in a mask-weighted manner to form a per-step \ac{UWB} feature representation.
% The exponential form is adopted to follow a smooth decrease in the influence of measurement with increasing age and making it consistent with gradual down-weighting of old/previous observations used in recursive estimation methods.

% Unlike a fixed forgetting factor, \ac{AoI} enters the model as an explicit input and is paired with a learned, anchor-specific decay constant \(\tau_a\), so stale measurements are attenuated in a data-driven manner.
% Anchor-level features are aggregated using the \ac{AoI}-decayed mask 
\begin{equation}
     \mathbf{h}^{\text{uwb}}_t =
     \frac{\sum_{a=1}^{N_A} \tilde{m}_{t,a} \mathbf{h}^{\text{uwb}}_{t,a}}
     {\sum_{a=1}^{N_A} \tilde{m}_{t,a} + \varepsilon}
\end{equation}
where $\varepsilon > 0$ is a small constant added for numerical stability.

To construct a scalar proxy reflecting \ac{UWB} measurements support and temporal freshness, {\proposalName} computes two complementary scalar quality indicators at each time step $t$.
The first indicator reflects the instantaneous measurement availability $q^{\mathrm{raw}}_t$ as
\begin{equation}
    q^{\mathrm{raw}}_t = \frac{1}{N_A} \sum_{a=1}^{N_A} m_{t,a},
\end{equation}
where $m_{t,a} \in \{0,1\}$ is the availability of the \ac{UWB} measurement from anchor $a$ at time $t$.
The second indicator incorporates temporal freshness by considering age-dependent reliability $q^{\mathrm{decay}}_t$ as
\begin{equation}
    q^{\mathrm{decay}}_t = \frac{1}{N_A} \sum_{a=1}^{N_A} \tilde{m}_{t,a},
\end{equation}
% where $\tilde{m}_{t,a} = m_{t,a}\exp(-\tau^{\mathrm{age}}_{t,a}/\tau_a)$ represents age-decayed availability and $\tau_a$ is a learned anchor-specific decay constant.

The final scalar \ac{UWB} quality score is obtained by combining both indicators
\begin{equation}
    q_t = \frac{1}{2} q^{\mathrm{raw}}_t + \frac{1}{2} q^{\mathrm{decay}}_t.
\end{equation}
The first term reflects how many anchors are currently visible, while the second reflects how recent their measurements are. Equal weighting is adopted to balance instantaneous availability and temporal freshness without introducing additional hyperparameters. The network subsequently learns how to exploit this reliability signal during training.
% To balance instantaneous measurement availability and its temporally decayed value, equal weighting is provided rather than using an additional tuning parameter.
% The network then learns to use this combined quality signal during training.
% The resulting quality sequence is normalized within each temporal window to the range $[0,1]$ before fusion to ensure numerical stability and comparability across windows. 

\subsubsection{Reliability-Aware Cross-Modal Fusion}
At each time step $t$, {\proposalName} constructs a fused feature representation by concatenating modality-specific embeddings and reliability cues $\mathbf{z}_t$ as
\begin{equation}
    \mathbf{z}_t = 
    \big[
    \mathbf{h}^{\mathrm{imu}}_t,\;
    \mathbf{h}^{\mathrm{uwb}}_t,\;
    q_t,\;
    q^{\text{prior}}
    \big],
\end{equation}
where $q^{\text{prior}}$ represents a time-invariant anchor-quality prior computed from training statistics and concatenated to $\mathbf{z}_t$ at each time step.
The sequence $\mathbf{z}_t$ is processed by an \ac{LSTM} model 
\begin{equation}
    \mathbf{h}^{\text{rnn}}_t = \text{LSTM}(\mathbf{z}_t).
\end{equation}

To dynamically balance inertial and ranging information, {\proposalName} employs a learned attention gate that adapts relative sensor contribution under varying \ac{UWB} availability, while maintaining stable motion estimation during sparse ranging conditions.
The gating input $\mathbf{g}_t$ is defined as
\begin{equation}
    \mathbf{g}_t = 
    \big[
    \mathbf{h}^{\mathrm{imu}}_t,\;
    \mathbf{h}^{\mathrm{uwb}}_t,\;
    q_t
    \big],
\end{equation}
% where, $q^{\mathrm{raw}}_t$ denotes the fraction of anchors providing measurements at time $t$.
% normalized by the total number of anchors $A$, as
% \begin{equation}
% n_t = \frac{1}{A}\sum_{a=1}^{A} m_{t,a}.
% \end{equation}
% The \ac{IMU} attention weight is computed via a multilayer perceptron 
The attention gate predicts a scalar \ac{IMU} weight $\alpha_t \in [0,1]$ using a multilayer perceptron followed by a sigmoid activation as
\begin{equation}
\alpha_t = \sigma\!\left(f_{\mathrm{att}}(\mathbf{g}_t)\right),
\end{equation}
where $\sigma(\cdot)$ denotes the sigmoid function.
The attended fusion feature is then computed as
\begin{equation}
\mathbf{h}^{\mathrm{att}}_t =
\alpha_t \mathbf{h}^{\mathrm{imu}}_t
+
(1-\alpha_t)\mathbf{h}^{\mathrm{uwb}}_t.
\end{equation}
This creates a convex combination of inertial and ranging features, allowing the network to emphasize \ac{IMU} information during \ac{UWB} degradation and rely more on ranging when reliable measurements are available.
% A safety constraint enforces minimum \ac{IMU} dominance thus ensuring stable behavior during complete \ac{UWB} outages.
A simple dominance constraint is applied to prevent degenerate behavior during complete \ac{UWB} outages and does not otherwise restrict the learned fusion behavior
\begin{equation}
    \alpha_t \ge \alpha_{\min} \quad \text{if } q^{\mathrm{raw}}_t =0,
\end{equation}
where $\alpha_{\min}$ is a predefined lower bound.

\subsubsection{Incremental Motion Prediction}
The final prediction head estimates incremental motion updates rather than absolute positions, improving stability under sparse measurements.
The predicted displacement $\Delta \hat{\mathbf{p}}_t$ at time $t$ is given as 
\begin{equation}
    \Delta \hat{\mathbf{p}}_t =
    s \cdot \tanh\!\left(
    \frac{f_{\Delta}\big([\mathbf{h}^{\mathrm{rnn}}_t, \mathbf{h}^{\mathrm{att}}_t]\big)}{s}
    \right),
\end{equation}
where, $s$ is a step-scale parameter derived from training statistics. 
The complete trajectory is constructed by cumulative integration of the predicted motion increments denoted by $\Delta\hat{\mathbf{P}}=\{\Delta\hat{\mathbf{p}}_t\}_{t=1}^{T}$.

\subsection{\DGAN{}: Overfitting Avoidance}

Given that our field campaign is limited to one trajectory, a key concern is that {\proposalName} may overfit to trajectory-specific \ac{UWB} error characteristics such as anchor-dependent biases, multipath-induced heavy tails, and bursty outage patterns rather than learning transferable sensor fusion behavior.
To probe this risk, we introduce a diffusion-based \ac{UWB} residual augmentation strategy referred as {\DGAN} that generates alternative, physically plausible measurement corruptions for the same underlying motion trajectory.
This enables a controlled stress test of the model’s robustness to measurement variability without altering the ground-truth motion.
Specifically, we train a conditional diffusion model on \ac{UWB} residuals, defined as the deviation between measured ranges and their geometric ground-truth values.
During {\proposalName} training, we occasionally perturb real residuals as follows:
\begin{enumerate}
    \item Compute real residuals
    \(
        \epsilon^{\text{real}}_{t,a} = d_{t,a} - r^{\text{geom}}_{t,a}
    \)
    for valid measurements, where $r^{\text{geom}}_{t,a}$ refers to the ground-truth geometric range computed using ground-truth positions.

    \item Sample synthetic residuals \(
        \epsilon^{\text{fake}}_{t,a}
    \) from the diffusion model, conditioned on motion and anchor visibility patterns.
    \item For a random subset (about \(10\%\)) of valid points, form a soft mixture
    \[
        \epsilon^{\text{mix}}_{t,a}
        = (1-\alpha_{\text{gan}})\,\epsilon^{\text{real}}_{t,a}
        + \alpha_{\text{gan}}\,\epsilon^{\text{fake}}_{t,a},
    \]
    where $\alpha_{\text{gan}} \in [0,1]$ controls the strength of the synthetic perturbation, and reconstruct augmented ranges
    \(
        d^{\text{aug}}_{t,a} = r^{\text{geom}}_{t,a} + \epsilon^{\text{mix}}_{t,a}.
    \)
\end{enumerate}

This augmentation strategy preserves the original sparsity pattern of the real \ac{UWB} data while injecting diversity into the noise structure.
Consequently, {\proposalName} observes a broader family of realistic \ac{UWB} degradations, reducing its sensitivity to any single noise fingerprint present in the original dataset.
This augmentation does not replace evaluation using additional trajectories; rather, it provides evidence that {\proposalName} does not rely on a specific measurement realization to achieve good performance.

% \begin{figure}[t]
% \centering
% \begin{subfigure}{0.48\linewidth}
%     \centering
%     \includegraphics[width=\linewidth]{figs/residual_cdf.pdf}
%     \caption{Cumulative distribution function (CDFs) of UWB measurement residuals obtained from real data and different GAN models.}
% \end{subfigure}
% \hfill
% \begin{subfigure}{0.48\linewidth}
%     \centering
%     \includegraphics[width=\linewidth]{figs/residual_absdev_stats.pdf}
%     \caption{Absolute deviation of key residual statistics (mean, median, $P95$, $P99$) from real-data values; smaller deviation indicates closer agreement.}
% \end{subfigure}
% \caption{Comparison of UWB residual distribution from different GAN models.}
% \label{fig:residual_eval}
% \end{figure}

% \begin{figure*}
%     \centering
%     \includegraphics[width=.8\textwidth]{figs/cdf_and_absdev_summary_combined.pdf}
%     \vspace{-.8em}
%     \caption{Comparison of UWB residual distribution from different GAN models. (Left) Cumulative distribution function (CDFs) of UWB measurement residuals obtained from real data and different GAN models. (Right) Absolute deviation of key residual statistics (mean, median, $P95$, $P99$) from real-data values; smaller deviation indicates closer agreement.}
%     \label{fig:gan_comparison}
%     \vspace{-.8em}
% \end{figure*}

\begin{figure}
    \centering
    \subfloat[Cumulative distribution function (CDFs) of UWB measurement residuals obtained from real data and different GAN models.]{%
        \includegraphics[width=.7\columnwidth]{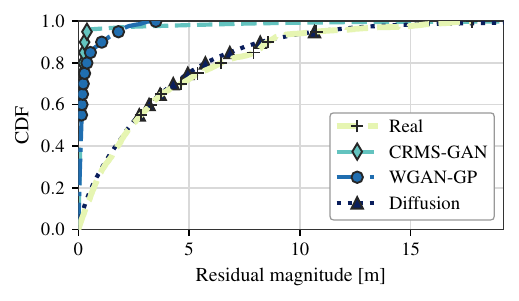}
        \label{fig:gan_cdf}
    }%
    \\
    \subfloat[Absolute deviation of key residual statistics (mean, median, $P95$, $P99$) from real-data values; smaller deviation indicates closer agreement.]{%
        \includegraphics[width=.7\columnwidth]{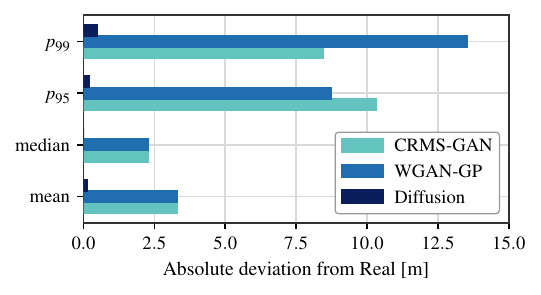}
        \label{fig:gan_err}
    }%
    \caption{Comparison of UWB residual distribution for different GAN models.}
    \vspace{-.8em}
\end{figure}

The proposed diffusion-based generator (DGAN) produces residual patterns that more accurately reflect the heavy-tailed, anchor-dependent nature of real \ac{UWB} noise compared to the two baseline generative models we evaluated for data augmentation (CRMS-GAN and WGAN-GP).
As shown in \cref{fig:gan_cdf}, the DGAN-generated residuals exhibit a CDF that more closely matches the real data with residual distribution that concentrates more probability mass near zero, hence a steeper rise of the CDF curve.
% This is quite evident from the CDF comparison  in \cref{fig:gan_comparison}.
% To match the residual distribution of the real dataset containing heavy-tailed behavior, DGAN produces the residual distribution that concentrates more probability mass near zero, hence a steeper rise of the CDF curve.
%
The deviation plot in \cref{fig:gan_err} further quantifies the absolute difference between model-generated and real residual statistics where the diffusion model shows the lowest deviation across evaluated residual statistics, particularly at high percentiles compared to the GAN baselines.
As a result, \ac{UWB} perturbations generated by DGAN better replicate real noise patterns, contributing to improved training robustness of {\proposalName} under sparsity patterns and \ac{UWB} degradation regimes compared to training without data augmentation.

\subsection{Loss Function and Training Strategy}

The training objective of {\proposalName} is defined as a composite robust regression loss that jointly promotes accurate short-term motion estimation, long-term motion consistency, and robust sensor fusion under sparse or degraded \ac{UWB} conditions and yields temporally stable attention behavior without requiring explicit attention regularization.
Training relies on Huber (Smooth L1) loss, which provides a balance between sensitivity to small errors with robustness to large deviations.
For a scalar error $e$, the Huber loss is 
\begin{equation}
\label{eq:huber_loss}
    \mathcal{L}_{\text{Huber}}(e) =
    \begin{cases}
    \frac{1}{2} e^{2}, & |e| \le \delta, \\[4pt]
    \delta \left(|e| - \frac{1}{2}\delta\right), & |e| > \delta,
    \end{cases}
\end{equation}
where $\delta$ denotes a transition threshold.
In our setting, $\mathcal{L}_{\text{Huber}}$ is applied element-wise to the 3D residuals between predicted and ground-truth incremental displacements and their cumulative positions, while a separate tail term operates on the Euclidean endpoint error.
The loss behaves quadratically for small residuals and linearly for large errors, providing robustness to large outliers and intermittent measurement dropouts. 

To avoid temporal leakage and preserve causality, data are split at the sequence level rather than at individual time steps.
Sliding windows of length $T$ are extracted from each trajectory, and exclusively assigned to training, validation, and testing set.
Normalization statistics $(\mu,\sigma)$ are computed using only the training set and reused unchanged for validation and test data.
Missing \ac{UWB} measurements within a window are causally filled using the hold of the last-observation-carried-forward strategy. So, the missing value of \ac{UWB} at $t$ is replaced by the most recent valid measurement at $t' < t$ to ensure causal consistency. No information from future timesteps $t' > t$ is used during preprocessing.

The network is trained end-to-end using mini-batch stochastic optimization with an adaptive learning rate.
Fixed-duration temporal segments sampled from continuous trajectories are employed for training.
To ensure stable convergence, the recurrent backbone is initialized through a brief warm-up stage and the age-aware decay parameters before activating auxiliary regularization terms such as reliability constraints and attention smoothness.
An optional diffusion-based generative augmentation module is used during training to assess robustness under measurement variability.
This module produces more realistic \ac{UWB} range residuals conditioned on motion platform and anchor visibility and is applied only to the training set.
Rather than replacing real measurements, the generative model injects small, physically plausible perturbations into valid \ac{UWB} observations, encouraging the fusion network to generalize beyond the limited set of observed \ac{UWB} patterns.
To avoid destabilizing early training, a curriculum strategy gradually increases the probability of augmented windows across training epochs, ensuring that the network first learns from clean data before being exposed to synthetic perturbations.
Model selection and early stopping are based on validation endpoint error, and all normalization parameters are derived solely from the training data to prevent information leakage.

\begin{figure}
    \centering
    \includegraphics[width=.8\linewidth]{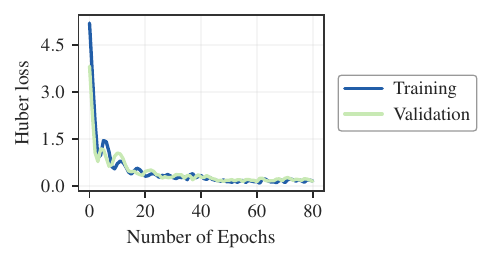}
    \vspace{-.8em}
    \caption{Training and validation endpoint loss as a function of training epochs. The close alignment of curves indicates stable optimization and minimal overfitting.}
    \label{fig:loss_curve}
    \vspace{-.8em}
\end{figure}

% \Cref{fig:loss_curve} shows the evolution of training and validation endpoint (final displacement error over a complete window) loss across epochs during model training.
\Cref{fig:loss_curve} shows the evolution of training and validation loss over epochs during model training, computed using Huber loss from \cref{eq:huber_loss}
The initial rapid drop in loss for early epochs corresponds to efficient learning of motion dynamics, and the subsequent plateau indicates convergence during early stopping.
The close agreement of the two curves suggests stable convergence with no signs of overfitting. This stability is further supported by stochastic residual augmentation, which empirically reduces sensitivity to trajectory-specific noise patterns.

\section{Performance Evaluation}\label{sec:performance_evaluation}

This section evaluates the performance of the proposed {\proposalName} under sparse and intermittent \ac{UWB} anchor observability. 

\begin{figure}
    \centering
    \includegraphics[width=.9\linewidth]{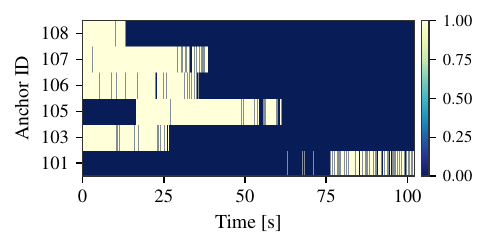}
    \vspace{-.8em}
    \caption{Temporal variation of UWB anchors illustrating intermittent anchor visibility along the experimental trajectory. Although six anchors are deployed, only a subset is visible at any time, with no more than four anchors simultaneously available.}
    \label{fig:anchors_vs_time}
    \vspace{-.8em}
\end{figure}

\Cref{fig:anchors_vs_time} displays the temporal availability of UWB anchors along the experimental trajectory. Multipath reflections, terrain occlusions, and varying \ac{LOS} conditions cause fluctuations in anchor visibility over time, leading to extended intervals when fewer than four anchors are available. Under such intermittent-observability conditions, standalone 3D trilateration is frequently infeasible, and the position estimates are often affected by high \ac{GDOP}.
This indicates the limitation of trilateration-based and loose-coupled based fusion methods, motivating tightly coupled ranging-inertial fusion.

The evaluation of the proposed model is performed on the test segment, and localization accuracy is examined using \ac{RMSE}, \ac{MAE}, and percentile errors ($P50$, $P95$, $P99$), capturing both average performance and tail robustness.
We evaluate {\proposalName} under two training configurations:
(i) the standard variant trained on real measurements, and (ii) a version trained with diffusion-based augmentation referred to as {\DGAN}.
Both variants use the same training and inference pipeline, DGAN version only differs at the training stage, where augmentation is applied and is not part of inference.
Performance is compared against the following baselines:
\begin{itemize}

    \item UWB-only Multilateration: Direct position estimation using available \ac{UWB} ranges.

    \item LC-AKF: A loose-coupling based Adaptive Kalman filtering method where \ac{UWB} trilaterated positions are fused with \ac{IMU} acceleration.

    \item LC-Bi-LSTM: A data-driven loose-coupling approach that fuses \ac{IMU} data with \ac{UWB}-derived positions using a bidirectional LSTM.

\end{itemize}

\subsection{Comparison of Evaluation Results}

\begin{table}
    % \vspace{-.8em}
    \centering
    \caption{Localization Error Statistics (meters)}
    \label{tab:localization_metrics}
    \begin{tabular}{lccccc}
        \toprule
        {Method} & {RMSE} & {MAE} & {$P50$} & {$P95$} & {$P99$} \\
        \midrule
        UWB-only          & 1.954 & 1.742 & 1.645 & 3.186 & 3.949 \\
        LC-AKF            & 1.701 & 1.485 & 1.369 & 3.372 & 3.638 \\
        LC-Bi-LSTM        & 0.312 & 0.279 & 0.389 & 0.416 & 0.439 \\
        \midrule
        {\proposalName}     & 0.136 & 0.110 & 0.086 & 0.276 & 0.383 \\
        \textbf{{\DGAN}}     & \textbf{0.129} & \textbf{0.106} & \textbf{0.082} & \textbf{0.259} & \textbf{0.355} \\
        \bottomrule
    \end{tabular}
    \vspace{-.8em}
\end{table}

\begin{figure}
    % \vspace{-.8em}
    \centering
    \includegraphics[width=.7\linewidth]{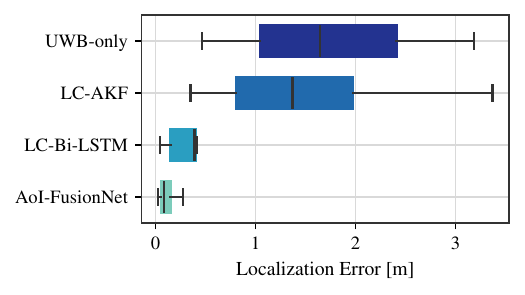}
    \vspace{-.8em}
    \caption{Distribution of absolute localization error for UWB-only, LC-AKF, LC-Bi-LSTM, and {\proposalName}.
    The boxes indicate the interquartile range (25th-75th percentiles), the central line denotes the median error, and the whiskers represent the 50th and 95th percentiles.}
    \label{fig:abs_error}
    \vspace{-.8em}
\end{figure}

\Cref{fig:abs_error} and \Cref{tab:localization_metrics} summarizes the localization error statistics across all methods.
Classical UWB-only multilateration and \ac{AKF}-based loose coupling exhibit large errors and heavy tails, primarily due to unreliable geometry and intermittent \ac{UWB} coverage.
The LC-Bi-LSTM baseline considerably improves performance by learning temporal motion patterns; however, it remains sensitive to \ac{UWB} outages because it relies on pre-computed \ac{UWB} positions requiring at least four anchors for reliable 3D localization.
{\proposalName} achieves the lowest RMSE and MAE, while also reducing tail errors ($P95$ and $P99$).
RMSE drops from \SI{0.31}{\meter} for LC-\ac{Bi-LSTM} to \SI{0.14}{\meter} for {\proposalName}.
These results indicate that tightly coupling raw \ac{UWB} ranges with \ac{IMU} data, rather than relying on intermediate trilateration, produces more stable trajectory estimates under sparse and noisy ranging conditions.
Incorporating diffusion-based augmentation further improves robustness, particularly in the upper error percentiles.
For example, $P99$ error is reduced from \SI{0.43}{\meter} LC-Bi-LSTM to \SI{0.38}{\meter} and \SI{0.35}{\meter} for {\proposalName} and {\DGAN} respectively.

\subsection{Error Distribution and Tail Robustness}

\begin{figure}
    \centering
    \includegraphics[width=.9\linewidth]{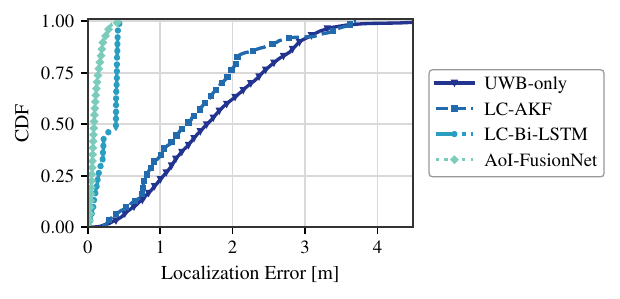}
    \vspace{-.8em}
    \caption{ \acp{CDF} of absolute localization error for UWB-only, LC-AKF, LC-Bi-LSTM, and {\proposalName} indicating that {\proposalName} achieves higher probability mass at lower error thresholds.}
    \label{fig:cdf}
    \vspace{-.8em}
\end{figure}

We compare the absolute localization error of {\proposalName} with multilateration and fusion baselines.
\Cref{fig:abs_error} shows that our proposed method exhibits the lowest median error and the narrowest interquartile range, indicating higher precision and greater consistency compared to the baselines.
UWB-only and LC-AKF show substantially large interquantile ranges and long upper tails with wider spreads, reflecting increased error and variability under intermittent anchor visibility.
Although LC-Bi-LSTM improves over traditional methods, its error distribution remains notably broader than {\proposalName}.
Overall, our proposed method achieves reduced error dispersion across the evaluated conditions.
\Cref{fig:cdf} illustrates the distribution and cumulative behavior of localization errors.
While multilateration and baseline fusion methods exhibit heavy tails, the distribution curve of {\proposalName} shows a steeper rise and reduced tail, highlighting improved robustness. 

\begin{figure}
    \centering
    \includegraphics[width=.9\linewidth]{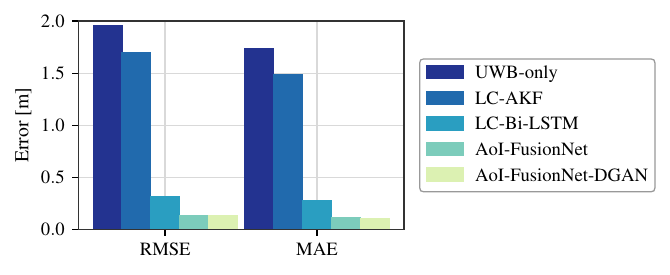}
    \vspace{-.8em}
    \caption{\ac{RMSE} and \ac{MAE} of 3D localization across all methods. {\proposalName} reduces RMSE from approximately 0.31 m (LC-Bi-LSTM) to about 0.14 m, while {\DGAN} further improves performance to about 0.13 m. Similar trends are observed for MAE, confirming consistent gains in both average and squared error measures.}
    \label{fig:bar_rmse_norm}
    \vspace{-.8em}
\end{figure}

\begin{figure}
    % \vspace{-.8em}
    \centering
    \includegraphics[width=.9\linewidth]{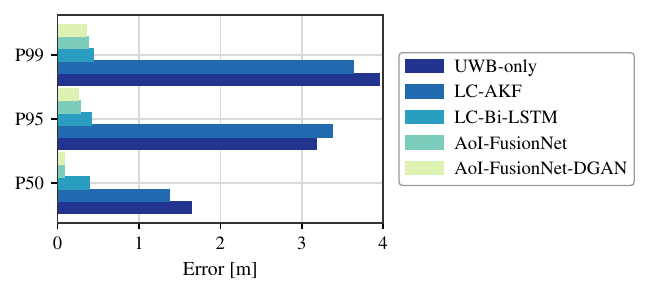}
    \vspace{-.8em}
    \caption{Median ($P50$), 95th percentile ($P95$), and 99th percentile ($P99$) 3D localization errors (Euclidean norm). {\proposalName} and {\DGAN} exhibit reduced tail errors compared to all baselines, indicating improved robustness under adverse sensing conditions.}
    \label{fig:percentile_bar}
    \vspace{-.8em}
\end{figure}

\Cref{fig:bar_rmse_norm,fig:percentile_bar} summarize quantitative comparisons of the average and tail localization errors across all methods. The DGAN variant of the proposed model is illustrated here through these metrics since it acts as a training-time augmentation mechanism rather than a separate inference model.
\Cref{fig:bar_rmse_norm} reports \ac{RMSE} and \ac{MAE}, showing that the proposed model and its diffusion-augmented version achieve lower mean errors than baselines.
In particular, {\DGAN} achieves the lowest \ac{RMSE}, indicating that diffusion-based augmentation has provided improvement without degrading the proposed model.
\Cref{fig:percentile_bar} further examines robustness through percentile-errors at $P50$, $P95$, and $P99$ reflecting tail behavior.
Baseline methods exhibit substantial degradation in high-percentile regimes displaying sensitivity to \ac{UWB} intermittent availability and measurement unreliability.
However, {\proposalName} and {\DGAN} maintain lower high-percentile errors, demonstrating resilience to degrading ranging conditions.

\subsection{Ablation of Attention (ATT) and \ac{AoI} Modules}

To assess the individual contribution of each architectural module independently, we have performed an ablation study summarized in \cref{tab:ablation}. 
The performance is reported in mean error metrics \ac{RMSE}, \ac{MAE}, and tail metrics ($P95$, $P99$).
The table shows that removing either the attention gate module or the \ac{AoI}-aware decay mechanism results in decreased mean localization and higher percentile errors. 
Overall, the primary impact of the \ac{AoI}-aware decay and attention gate in our proposed {\proposalName} model is depicted in tail metrics, rather only the mean errors, suggesting that these modules contribute most to prevent severe localization failures under intermittent measurement availability.

\begin{table}[b]
    \vspace{-.8em}
    \centering
    \caption{Ablation results (localization errors in meters) for attention (ATT) and AoI modules using raw UWB ranges and IMU data.}
    \label{tab:ablation}
    \begin{tabular}{cccccc}
        \toprule
        ATT & AoI & RMSE & MAE & $P95$ & $P99$ \\
        \midrule
        \cmark & \cmark & 0.136 & 0.110 & 0.276 & 0.383 \\
        \cmark & \xmark & 0.165 & 0.140 & 0.320 & 0.413 \\
        \xmark & \cmark & 0.223 & 0.185 & 0.438 & 0.524 \\
        \xmark & \xmark & 0.189 & 0.160 & 0.342 & 0.465 \\
        \bottomrule
    \end{tabular}
    % \vspace{-.8em}
\end{table}
% \begin{table}
%     \vspace{-.8em}
%     \centering
%     \caption{Ablation results (localization errors in meters) for attention (ATT) and AoI modules using raw UWB ranges and IMU data.}
%     \label{tab:ablation}
%     \begin{tabular}{cccccccc}
%         \toprule
%         ATT & AoI & RMSE & MAE & $P95$ & $P99$ & RMSE$\ge2$ & $P95$$\ge2$ \\
%         \midrule
%         \cmark & \cmark & 0.136 & 0.111 & 0.276 & 0.383 & 0.145 & 0.254 \\
%         \cmark & \xmark & 0.165 & 0.140 & 0.320 & 0.413 & 0.201 & 0.353\\
%         \xmark & \cmark & 0.223 & 0.185 & 0.438 & 0.524 & 0.166 & 0.276 \\
%         \xmark & \xmark & 0.189 & 0.160 & 0.342 & 0.465 & 0.199 & 0.340 \\
%         \bottomrule
%     \end{tabular}
%     % \vspace{-.8em}
% \end{table}

\begin{figure}
    \centering
    \subfloat[Mean IMU attention weight conditioned by anchor regime]{%
        \includegraphics[width=.7\columnwidth]{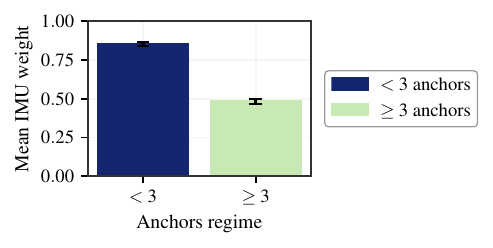}
        \label{fig:gate_mean}
    }%
    \\
    \subfloat[Temporal evolution of fusion gate]{%
        \includegraphics[width=\columnwidth]{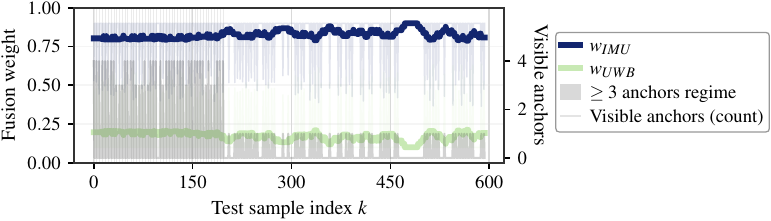}
        \label{fig:gate_attention}
    }%
    \caption{Behavior of the learned fusion gate.
    (a) Under sparse UWB conditions ($<3$ visible anchors), the estimator operates primarily in the inertial propagation regime, resulting in high $w_{\mathrm{IMU}}$, while greater anchor availability leads to reduced IMU weighting.
    (b) Time-resolved attention weights $w_{\mathrm{IMU}}$, $w_{\mathrm{UWB}}$ along with the number of visible anchors over the test sequence.  The predominance of $w_{\mathrm{IMU}}$ reflects the intermittent and limited ranging geometry in this scenario, with \ac{UWB} acting primarily as a corrective constraint.}
    \vspace{-.8em}
\end{figure}
% \todo{please crop these figures (remove white borders), this will make them a little bitter and the text better readable}

Finally, we examine the behavior of the learned fusion gate.
As shown in \cref{fig:gate_mean}, the model assigns higher weight to \ac{IMU} data under sparse condition ($<3$ visible \ac{UWB} anchors), depicting greater reliance on inertial propagation when anchor geometry is unreliable and less weight when sufficient anchors are visible.
\Cref{fig:gate_attention} shows the temporal evolution of the fusion gate over the test sequence.
The $w_{\mathrm{IMU}}$ and $w_{\mathrm{UWB}}$ refer to the learned fusion weights assigned to inertial and ranging measurement at each time step.
The vertical spikes represent the instantaneous number of visible \ac{UWB} anchors providing a direct illustration of \ac{UWB} measurement availability over time.
The evolution of weights remains smooth with changes in anchor availability, confirming gradual adaptation rather than rapid temporal switching.
This behavior demonstrates that the gating mechanism responds in a physically meaningful manner to measurement availability while preserving stable fusion dynamics.

% \subsection{COMPARISON OF COMPUTATIONAL RUNTIME}

\subsection{Discussion}

The results of the proposed {\proposalName} algorithm exhibit consistent localization enhancement as compared to classical \ac{UWB} multilateration and loose coupling fusion baselines such as Bayesian filtering and deep learning in a challenging \ac{GNSS}-denied outdoor scenario. 
The improvements are notably evident in high-percentile error metrics, highlighting its robustness under intermittent \ac{UWB} observability instead of gradual degradation of measurement quality for the evaluated dataset.
The empirical analysis of anchor availability and \ac{AoI} suggests that the informative \ac{UWB} mostly appear in short intermittent bursts, and these measurements are mostly fresh when available.
In this regime, the proposed \ac{AoI}-aware decay mechanism helps the model to reduce the influence of stale measurements when fresh data is unavailable.
%
% Here, the proposed \ac{AoI}-aware decay mechanism functions mainly as a freshness prior complementing availability-based fusion, attenuating stale measurements, and avoiding sudden weight changes.

The learned fusion gate shows smooth and adaptive weighting of inertial and \ac{UWB} cues, exploiting informative anchor geometry when available.
In this way, the model does not enforce position estimates when \ac{UWB} measurements are insufficient.
Instead, the model relies on learned motion from \ac{IMU} to track smoothly and corrects drift whenever reliable \ac{UWB} measurements become available.
An \ac{LSTM} backbone is employed for its stable training and lower data requirements than other parameter-intensive models. Due to the single-layer \ac{LSTM} design and anchor-wise shared encoders, the model is computationally lightweight, making it suitable for offline post-processing.
To mitigate the risk of overfitting to a single motion pattern, a diffusion-based residual augmentation technique has been used to produce physically plausible \ac{UWB} noise patterns for the same underlying motion trajectory.
This augmentation does not substitute for an evaluation of additional trajectories; however, it serves as a stress test showing that the proposed model does not rely on a specific noise fingerprint. % to perform well.

Although the proposed model shows improved performance in the evaluated scenario, there are some limitations as well.
First, the data comes from an environment with fixed anchor deployment, leaving room for cross-environment generalization.
Secondly, the model does not explicitly distinguish between \ac{LOS} and \ac{NLOS} effects but learns to reduce the impact of unreliable measurements using temporal patterns and \ac{AoI}.
Despite these limitations, our tightly coupled, \ac{AoI}-aware fusion model effectively handles intermittent sensing conditions without heuristic filtering.

\section{Conclusion}\label{sec:conclusion}

In this work, we proposed an adaptive tightly-coupled \ac{UWB}-\ac{IMU} fusion method, {\proposalName}, to perform 3D localization in \ac{GNSS}-denied environments under sparse and intermittent rangings.
The method integrates temporal motion dynamics, age-aware reliability modeling of \ac{UWB} measurements using \ac{AoI}, and adaptive sensor weighting through learned attention gate to achieve lower localization error, particularly in high-percentile error metrics, than multilateration, Kalman filtering, and deep learning based loose coupling fusion methods.
The evaluation conducted during offline post-processing of real-world data collected in a challenging alpine environment reveals that the proposed model remains stable and accurate when informative \ac{UWB} geometry is available only intermittently.
The ablation results further demonstrate that \ac{AoI}-aware decay and attention-based fusion provide complementary roles, with the largest gains observed in high-percentile error metrics.

Future work will focus on the extension of this framework, including multi-node and cooperative localization, cross-environment validation, and \ac{CIR}-based explicit \ac{LOS}/\ac{NLOS} modeling through joint classification to enhance robustness under severe multipath conditions.
Together, these directions pave the way for a more general and transferable learning-based tracking model suitable for diverse \ac{GNSS}-denied scenarios.

% \bibliographystyle{ACM-Reference-Format}
% \bibliography{references}
\printbibliography

\end{document}